\documentclass[sigconf]{acmart}
\AtBeginDocument{%
  }

\copyrightyear{2023} 
\acmYear{2023} 
\setcopyright{acmlicensed}\acmConference[ICPP 2023]{52nd International Conference on Parallel Processing}{August 7--10, 2023}{Salt Lake City, UT, USA}
\acmBooktitle{52nd International Conference on Parallel Processing (ICPP 2023), August 7--10, 2023, Salt Lake City, UT, USA}
\acmPrice{15.00}
\acmDOI{10.1145/nnnnnnn.nnnnnnn}
\acmISBN{978-x-xxxx-xxxx-x/YY/MM}

\usepackage{acmart-taps}
\usepackage{graphicx}
\usepackage{subcaption}

\usepackage{listings}
\usepackage{fancyvrb}
\DefineVerbatimEnvironment{myVerbatim}{Verbatim}
{
  frame=lines,
  framesep=2mm,
    fontsize=\scriptsize,
}

\usepackage[ruled,vlined,linesnumbered]{algorithm2e}
\RestyleAlgo{ruled}
\usepackage{enumitem}

\usepackage{lipsum} 
\usepackage{titlesec}

\begin{document}

\title{FaST-GShare: Enabling Efficient Spatio-Temporal GPU Sharing in Serverless Computing for Deep Learning Inference}
% efficient deep learning inference in serverless computing using   Spatio-Temporal GPU Sharing

% \affiliation{%

\author{Jianfeng Gu$^*$\textdagger, \, Yichao Zhu$^*$, \, Puxuan Wang, \, Mohak Chadha, \, Michael Gerndt}
\affiliation{%
  \institution{Chair of Computer Architecture \& Parallel Systems, Technical University of Munich}
  \city{Munich}
  \country{Germany}}
\email{ {jianfeng.gu, yukio.zhu, puxuan.wang, mohak.chadha}@tum.de, gerndt@in.tum.de}

\thanks{$^*$Equal contribution. \textdagger Corresponding Author.}

\renewcommand{\shortauthors}{Jianfeng Gu et al.}

\begin{abstract}
Serverless computing (FaaS) has been extensively utilized for deep learning (DL) inference due to the ease of deployment and pay-per-use benefits. However, existing FaaS platforms utilize GPUs in a coarse manner for DL inferences, without taking into account spatio-temporal resource multiplexing and isolation, which results in severe GPU under-utilization, high usage expenses, and SLO (Service Level Objectives) violation. There is an imperative need to enable an efficient and SLO-aware GPU-sharing mechanism in serverless computing to facilitate cost-effective DL inferences. In this paper, we propose \textbf{FaST-GShare}, an efficient \textit{\textbf{Fa}aS-oriented \textbf{S}patio-\textbf{T}emporal \textbf{G}PU \textbf{Sharing}} architecture for deep learning inferences. In the architecture, we introduce the FaST-Manager to limit and isolate spatio-temporal resources for GPU multiplexing. In order to realize function performance, the automatic and flexible FaST-Profiler is proposed to profile function throughput under various resource allocations. Based on the profiling data and the isolation mechanism, we introduce the FaST-Scheduler with heuristic auto-scaling and efficient resource allocation to guarantee function SLOs. Meanwhile, FaST-Scheduler schedules function with efficient GPU node selection to maximize GPU usage. Furthermore, model sharing is exploited to mitigate memory contention. Our prototype implementation on the OpenFaaS platform and experiments on MLPerf-based benchmark prove that FaST-GShare can ensure resource isolation and function SLOs. Compared to the time sharing mechanism, FaST-GShare can improve throughput by 3.15x, GPU utilization by 1.34x, and SM (Streaming Multiprocessor) occupancy by 3.13x on average.

\vspace{-2mm}

\end{abstract}

\ccsdesc[500]{Computer systems organization~Cloud Computing}
\ccsdesc[300]{Computer systems organization~Serverless computing}
\ccsdesc{Computer systems organization~Deep learning}
\vspace{-4mm}

% \keywords{datasets, neural networks, gaze detection, text tagging}
% \keywords{Serverless computing, GPU sharing, deep leanrning inference, scheduling}



\maketitle
\vspace{-2mm}
\section{Introduction}
Serverless Computing, also known as Function-as-a-Service (FaaS), is emerging as a new paradigm \cite{jonas2019cloud} for the next generation of cloud-native computing due to the ease of deployment, high scalability, and cost-effective pay-per-use benefits. With serverless, users can build and run applications without considering the servers. Instead, cloud providers are responsible for allocating resources on demand. Serverless functions are designed to be event-driven and fine-grained so that they can scale elastically to accommodate workload changes, enabling resource sharing among clients. Due to these advantages, the adoption of serverless computing is increasing rapidly in various fields, including video analytics \cite{lixiang_socc_2018, Llama_socc_2021}, DAG workload processing \cite{Mahgoub_atc_2021}, linear algebra \cite{Vaishaal2020}, ML training and inferences \cite{jananie_icpp_2021, andreas_bigdata_2021, tetris2022, fedlessscan2022}, and etc.
% ahsan_vldb_2022
% Mahgoub_osdi_2022
% Mahgoub_osdi_2022
% fouladi_nsdi_2017
% cirrus_2019_socc
% ahsan_2020_sc

\vspace{-2mm}
\begin{figure}[htbp]
  \centering
  \begin{subfigure}[b]{0.23\textwidth}
    \includegraphics[width=\textwidth]{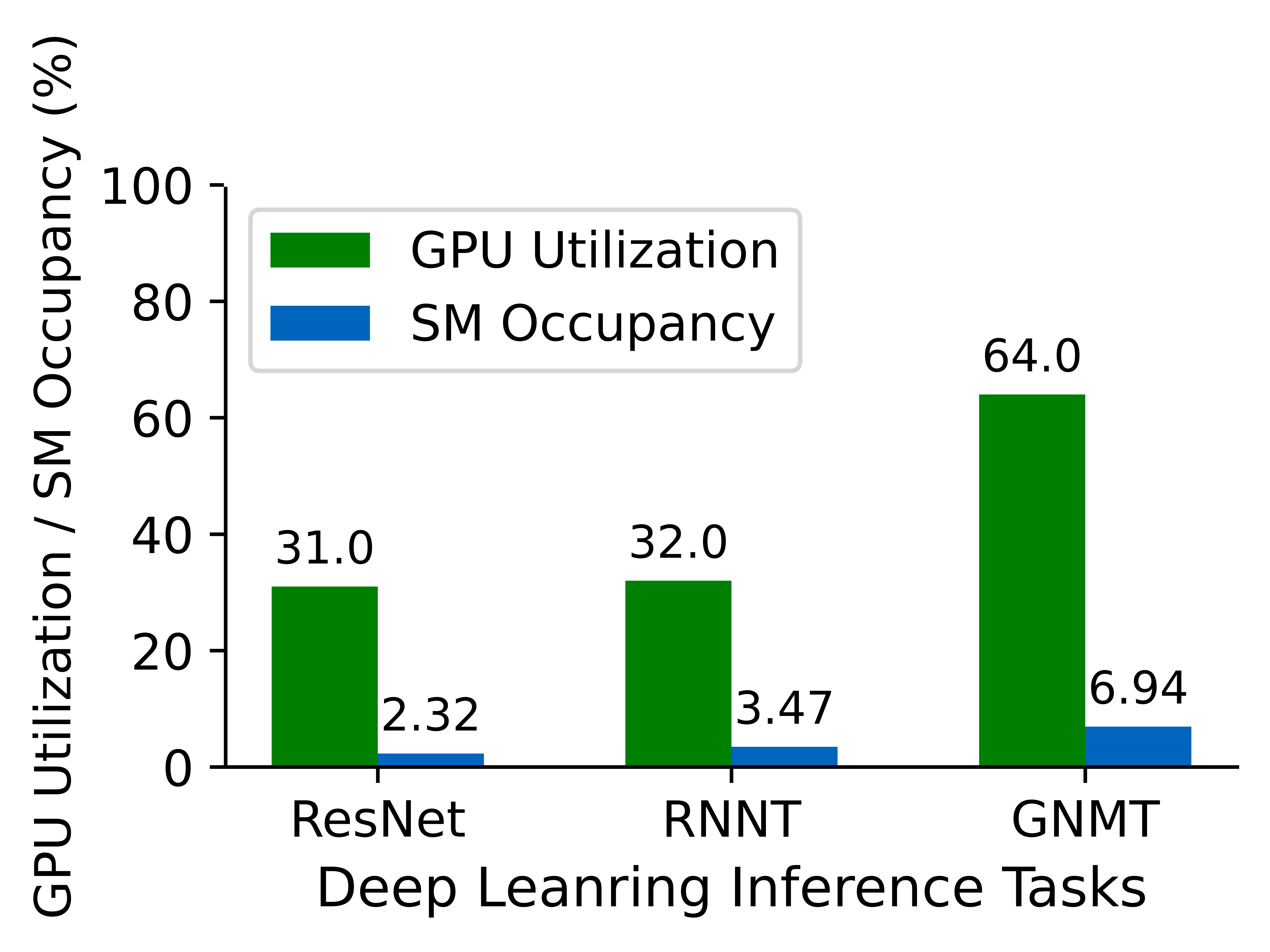}
    \vspace{-5mm}
    \caption{Kubernetes Device Plugin \cite{device_plugin}}
    \label{fig:device_plugin}
  \end{subfigure}
%   \quad % or any other spacing command you like
  \begin{subfigure}[b]{0.22\textwidth}
    \includegraphics[width=\textwidth]{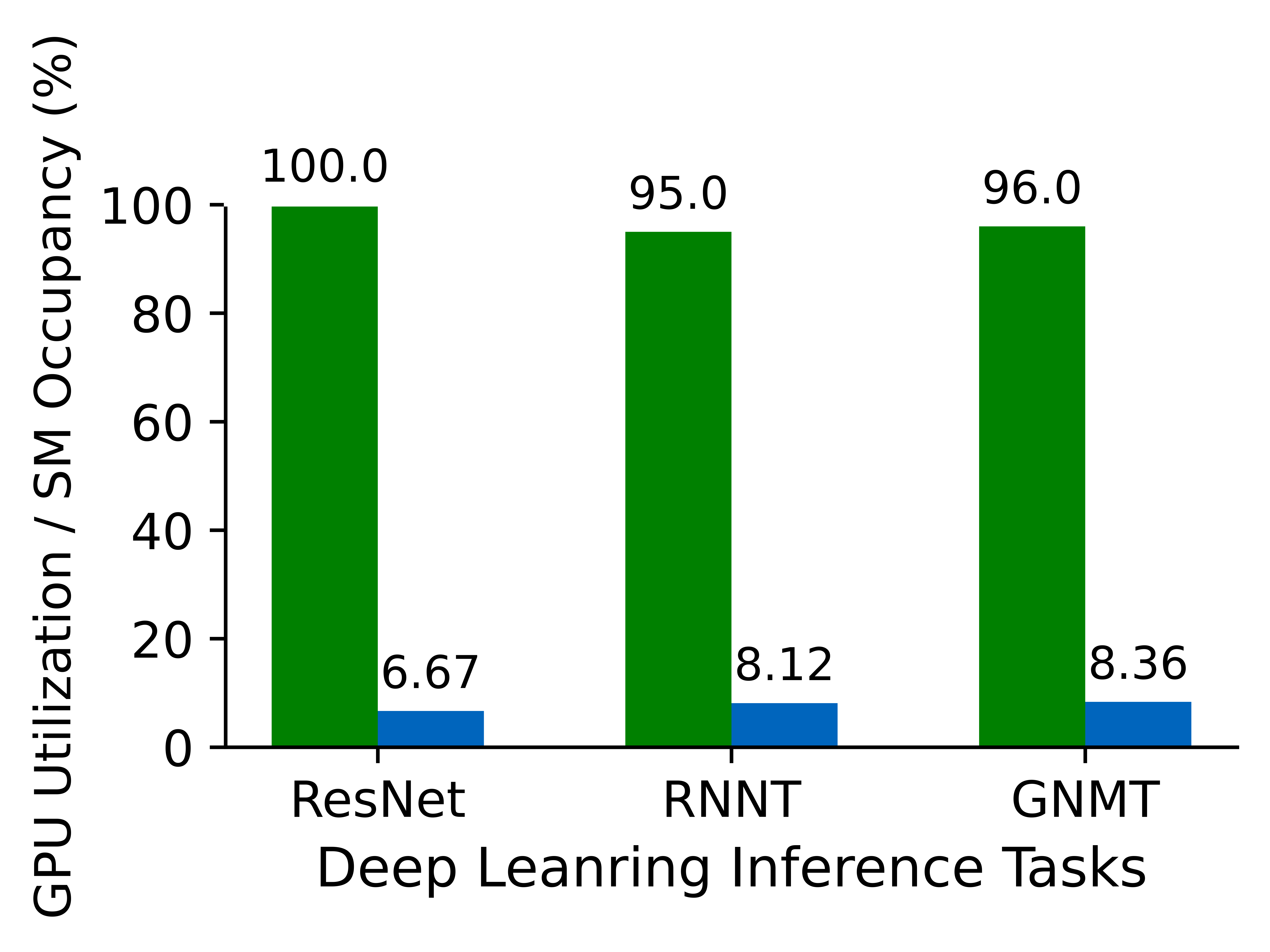}
    \vspace{-5mm}
    \caption{Time sharing \cite{kubeshare_hpdc_2020}.}
    \label{fig:time_sharing}
  \end{subfigure}
  \vspace{-3mm}
  \caption{GPU utilization and SM occupancy when using device plugin and timing sharing under extreme workload. }
  \label{fig:intro_gpu_utilization}
  \vspace{-2mm}
\end{figure}

FaaS platforms \cite{openfaas,knative,openwhisk} leverage Kubernetes to deploy and manage functions due to its powerful container orchestration capability. To facilitate the GPU usage in Kubernetes, NVIDIA offers device plugin \cite{device_plugin} to manage GPUs across multiples nodes. However, the plugin is typically designed for deep learning training by assigning an entire GPU device to a single container, which results in severe resource under-utilization in deep learning inferences, especially on modern data center GPUs with massive compute units, eg. NVIDIA V100 Tesla GPU (80 SM (Streaming Multiprocessor) units) and A100 HGX GPU (108 SM units). As shown in Figure \ref{fig:device_plugin}, even under extreme inference workloads, the GPU utilization remains very low. Although the Time-Slicing GPUs for Kubernetes \cite{time_slicing_gpu} was introduced by NVIDIA to share a GPU, it just enables the fair sharing and does not guarantee a pod to receive a proportional amount of GPU compute power. The lack of resource isolation will bring unpredictable interference among FaaS functions and fail to meet service level objectives (SLOs). Hence, it is desirable to design a both efficient and SLO-aware GPU sharing mechanism in serverless computing for deep learning inferences.

% \cite{gslice_socc_2020}

However, enabling efficient and SLO-aware GPU sharing in serverless computing is not a trivial problem. First, we find that current sharing mechanisms are not fine-grained and isolated enough to conform to the cost-effectiveness and intrinsic fine-grained design of FaaS functions. Previous works \cite{kubeshare_hpdc_2020, time_slicing_gpu, liu_icpp_2022} introduced time sharing policies for Kubernetes and container clouds. However, as shown in Figure \ref{fig:time_sharing}, the GPU utilization of running inference instances with timing sharing exceeds 95\%, but the SM occupancy only reaches less than 10\%, which will exacerbate with more powerful GPUs. Because the CUDA kernels of most DNN models cannot drain all GPU SMs at any given time.  Meanwhile, although recent spatial sharing policies \cite{gslice_socc_2020,mlsys_2022_junguk, choi_atc_2022} based on MPS (Multi-Process Service) \cite{mps_introduction} for specific ML inference systems improved SM utilization, they are so focused on performance that isolation or instance scalability is overlooked. Therefore, how to coordinate GPU spatial and temporal multiplexing to ensure both efficiency and isolation becomes the key to achieving efficient FaaS inferences. 

Second, with the spatio-temporal GPU sharing mechanism, how to allocate the appropriate amount of resources for each function and schedule functions across GPU nodes to ensure its SLO and throughput while improving GPU resource usage poses a challenge. To ensure the SLO latency, previous specific ML inference systems \cite{gslice_socc_2020, choi_atc_2022} allocate an initial GPU partition to the inference server and then dynamically re-provision resources according to workload variations. Unfortunately, their resource-based scaling mechanism within a single instance is not compatible with the inherent instance-based scaling used in serverless computing. 

Third, deep learning inference tasks are known to require a significant amount of GPU memory for deploying frameworks and parameter models. However, the fine-grained GPU sharing mechanism enables the deployment of more FaaS functions in a GPU, which can lead to increased contention for GPU memory. 

To address these challenges, we propose the \textit{\textbf{Fa}}aS-oriented \textbf{S}patio-\textbf{T}emporal \textbf{G}PU \textbf{Sharing} (\textbf{FaST-GShare}) architecture. FaST-GShare aims to enable efficient spatio-temporal GPU sharing in serverless computing for DL inferences, which can maximize FaaS function throughput, GPU utilization, and SM occupancy while ensuring function SLOs. In FaST-GShare, we propose \textbf{FaST-Manager} to coordinate the spatial and temporal GPU resource sharing by limiting and isolating resources in both two dimensions (2D) with frontend-backend architecture. FaST-manager adopts the mechanism of controlling spatial multiplexing through multiple concurrent time tokens to achieve the spatio-temporal limitation collaboration. 

To allocate the appropriate amount of resources for functions, first, we introduce the automatic and flexible \textbf{FaST-Profiler} to profile the function throughput and latency under various spatio-temporal GPU resource allocations. Based on the profiling information, \textbf{FaST-Scheduler} selects the most efficient spatial and temporal resource configurations and performs the auto-scaling for function instances through the Heuristic Scaling Algorithm. Meanwhile, FaST-Scheduler schedules functions with the Maximal Rectangles Algorithm that selects efficient GPU nodes and dynamically reorganizes GPU resources for functions to maximize GPU usage.

To address the problem of memory contention, we propose the \textbf{model sharing} mechanism where multiple function instances can share one copy of models in a memory-efficient way.

In a nutshell, the contributions are summarized as follows:
\begin{itemize}[topsep=0pt, parsep=0pt]
\item We propose FaST-GShare, an efficient spatio-temporal GPU sharing architecture in serverless computing for deep learning inference to maximize function throughput, GPU utilization, and SM occupancy while ensuring function SLO.
\item  We design a spatio-temporal GPU sharing mechanism, FaST-Manager, that can limit and isolate GPU resources with any temporal and spatial granularity. 
\item  We propose FaST-Profiler and FaST-Scheduler with the Heuristic Scaling Algorithm and the Maximum Rectangle Algorithm to achieve appropriate resource allocation, auto-scaling, and efficient function scheduling for FaaS DL inference. We design the model sharing mechanism to achieve up to a 55.6\% reduction in function memory footprint.
\item We implement FaST-GShare architecture on the OpenFaaS platform. The experiments on MLPerf-based \cite{mlperf_benkmarks} benchmark demonstrate that FaST-GShare can achieve effective resource isolation and ensure function SLOs. Compared to the time sharing mechanism, FaST-GShare can improve throughput by 3.15x, GPU utilization by 1.34x, and SM occupancy by 3.13x on average.
\end{itemize}

\vspace{-7pt}
\section{Background and Related Work}
\subsection{Serverless Computing}

Serverless computing is a cloud computing model that allows developers to deploy and run code without managing the underlying infrastructure. Function is a self-contained unit of code, like codes of a neural network model written with PyTorch, that can be executed in container and response to an event. It is cost-effective, scalable, and agile because users only pay for the resources used by their functions. Due to these advantages, cloud providers have introduced their serverless platforms ( Amazon Lambda \cite{aws_lambda}, Microsoft Azure Functions \cite{azure_functions}, and Google Cloud Functions \cite{google_functions} ). Meanwhile, several open-source serverless frameworks ( OpenFaaS \cite{openfaas}, Knative \cite{knative}, OpenWhisk \cite{openwhisk}, and etc. ) have been offered. However, the publicly available severless offerings from most cloud providers only allow CPU access currently and do not natively support accelerators, such as GPUs and TPUs. Indeed, recent works \cite{ tetris2022,ahsan_vldb_2022} mainly utilize CPU to perform machine learning inferences.

% Mahgoub_osdi_2022

\vspace{-5pt}
\subsection{GPU Device Plugin for Kubernetes}
Most FaaS platforms \cite{openfaas,knative,openwhisk} leverage Kubernetes to deploy and manage functions due to its powerful container orchestration capability. Kubernetes is a control plane-node architecture. The control plane manages the entire cluster, while the worker nodes are responsible for running pods. A pod is the smallest deployable unit that can be scheduled across nodes. The GPU device plugin \cite{device_plugin} in Kubernetes communicates with the NVIDIA device driver to discover and report the available GPUs on a node. It exposes this information to the Kubernetes control plane, which can then schedule pods with GPU-accelerated workloads onto nodes with available GPUs. The plugin also handles device assignment and lifecycle management for GPU workloads, ensuring that they have exclusive access to the assigned GPUs. However, the plugin is typically designed for deep learning training by assigning an entire GPU device to a single container, which results in severe resource under-utilization in deep learning inferences.

Recently, some timing sharing implementations based on device plugins have been proposed. NVIDIA offers Time-Slicing GPUs in Kubernetes \cite{time_slicing_gpu} to allow workloads scheduled on oversubscribed GPUs to interleave with one another. But it does not guarantee that the pod receives access to a proportional amount of GPU compute power. vGPU-scheduler \cite{k8s_vGPU} was proposed to allocate a portion of GPU to pods through time sharing. However, the method primarily focused on GPU memory limitation.

\vspace{-2mm}
\subsection{Fine-grained GPU Sharing}
There are two primary approaches for achieving fine-grained GPU sharing, namely temporal sharing and spatial sharing. Gaia-GPU \cite{gu2018gaiagpu} and Gemini \cite{chen2021gemini} are proposed to enable temporal GPU resource sharing and isolation. They both used the mechanism of intercepting CUDA driver API to impose time limits on GPU. KubeShare \cite{kubeshare_hpdc_2020} extended time sharing of GPU based on Gemini to the Kubernetes through vGPU abstraction. The paper utilized a similar interception mechanism to enable time sharing on GPU.

As for spatial sharing, NVIDIA’s latest Ampere GPU launched Multi-Instance GPU (MIG) to partition a GPU to multiple vGPU instances. However, this hardware-based partitioning is limited to only seven pre-defined resource configurations. GSLICE \cite{gslice_socc_2020} proposed a dynamically adaptive resource allocation to improve DL inference throughput. Choi et al. \cite{choi_atc_2022} introduced the mechanism of dynamically scaling spatio-temporal resources to ensure inference SLOs. These resource-based scaling mechanism within a single instance is not compatible with the inherent instance-based scaling used in serverless computing. NVIDIA offers MPS to enable CUDA kernels from different processes to be processed concurrently on the same GPU, which can significantly improve GPU performance. Junguk et al. \cite{mlsys_2022_junguk} utilized MPS to improve inference throughput, but without considering isolation. The paper utilized MPS to design the spatial sharing mechanism. It should be noted that the architecture proposed in this paper is compatible with any MPS-supported GPU, including both Volta-GPUs and Ampere/Hopper GPUs with MIG mechanism enabled, where multiple MPS clients can run simultaneously on each MIG vGPU instance.

\vspace{-5pt}
\section{Architecture}
\vspace{-2pt}
\subsection{Architecture Overview}

\aptLtoX[graphic=no,type=env]{\begin{figure}
  \centering
    \includegraphics[width=0.55\textwidth]{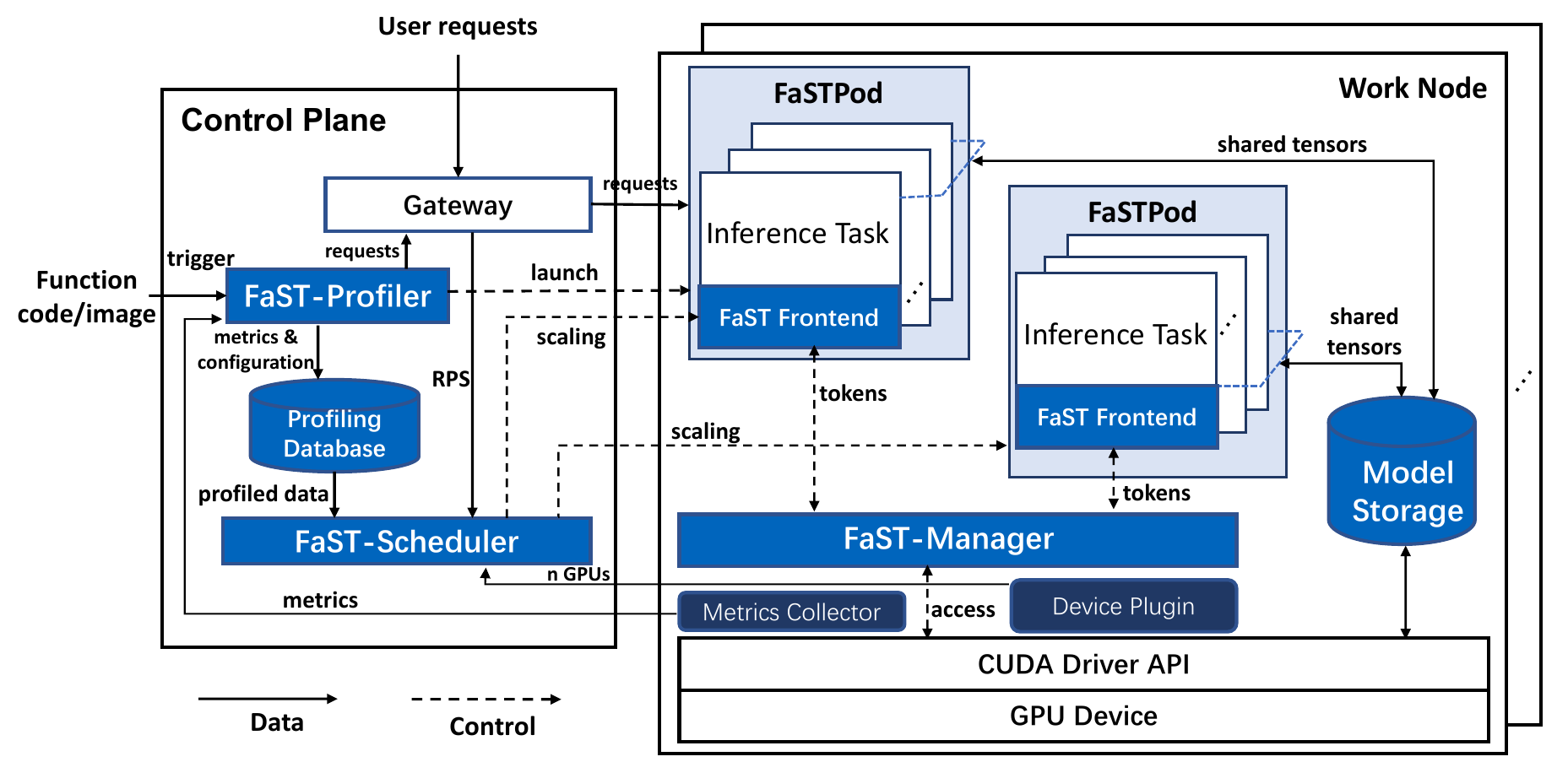}
    \caption{The architecture of FaST-GShare.}
    \label{fig:architecture}
\end{figure}\begin{figure}
    \centering
    \includegraphics[width=0.50\textwidth]{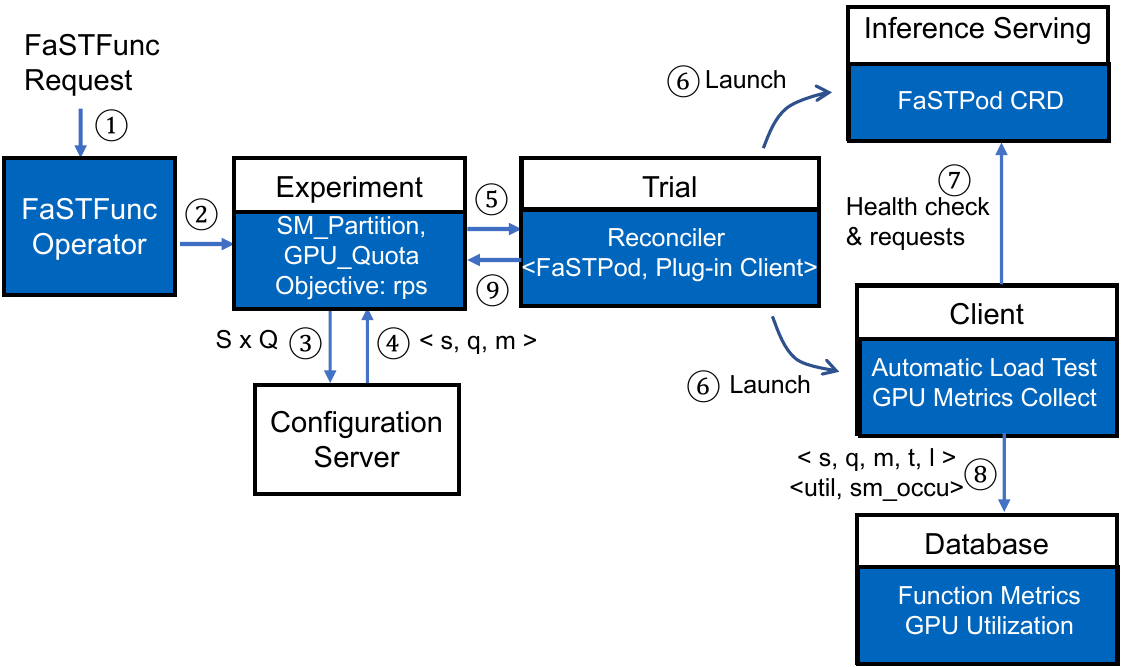}
    \caption{The workflow of the FaST-Profiler.}
    \label{fig:profiler}
\end{figure}}{\begin{figure*}[thbp]
  \centering
  \begin{minipage}[t]{0.55\textwidth}
    \centering
    \includegraphics[width=\textwidth]{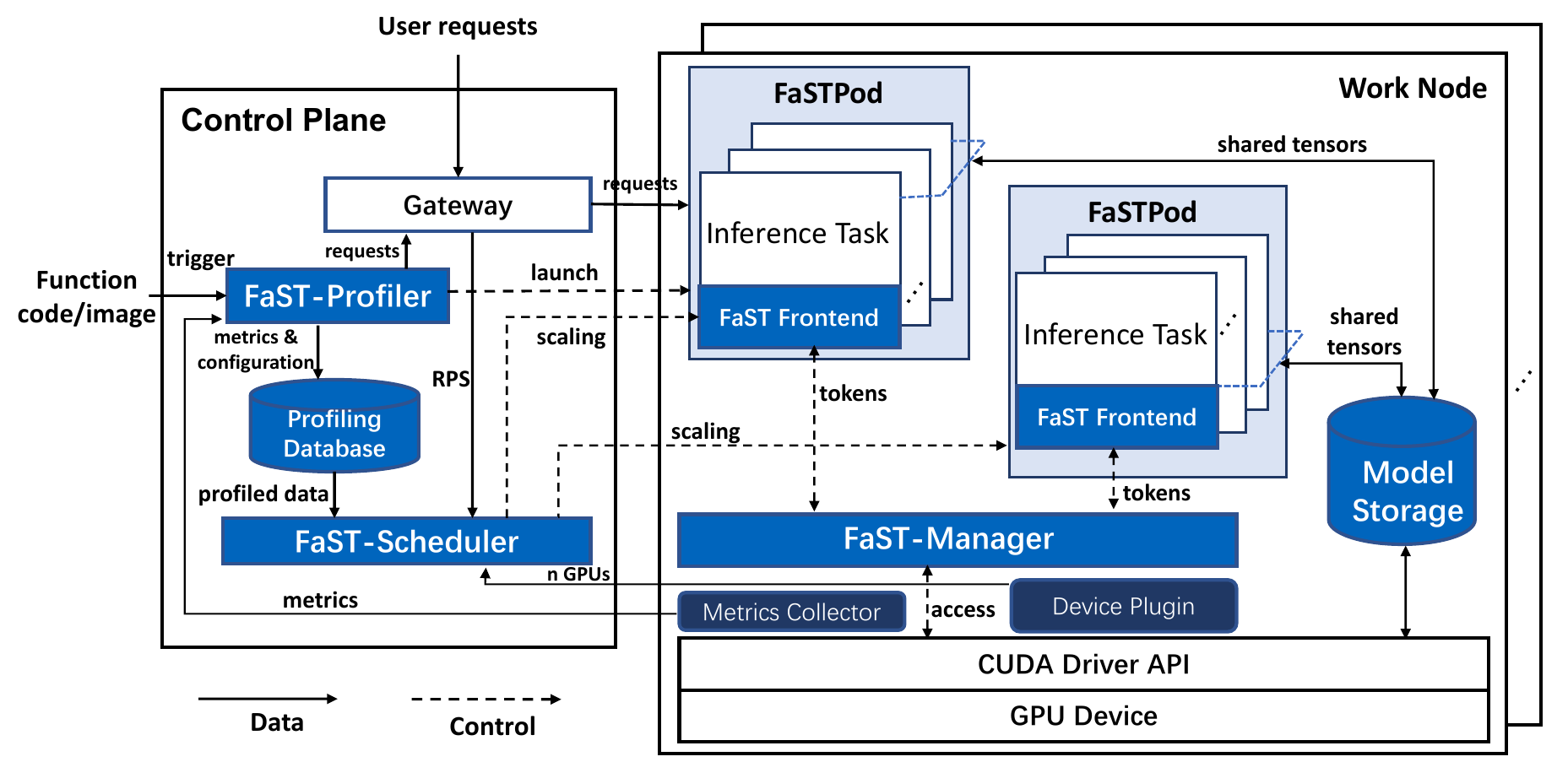}
    \caption{The architecture of FaST-GShare.}
    \label{fig:architecture}
  \end{minipage}
  \hfill
  \begin{minipage}[t]{0.44\textwidth}
    \centering
    \includegraphics[width=\textwidth]{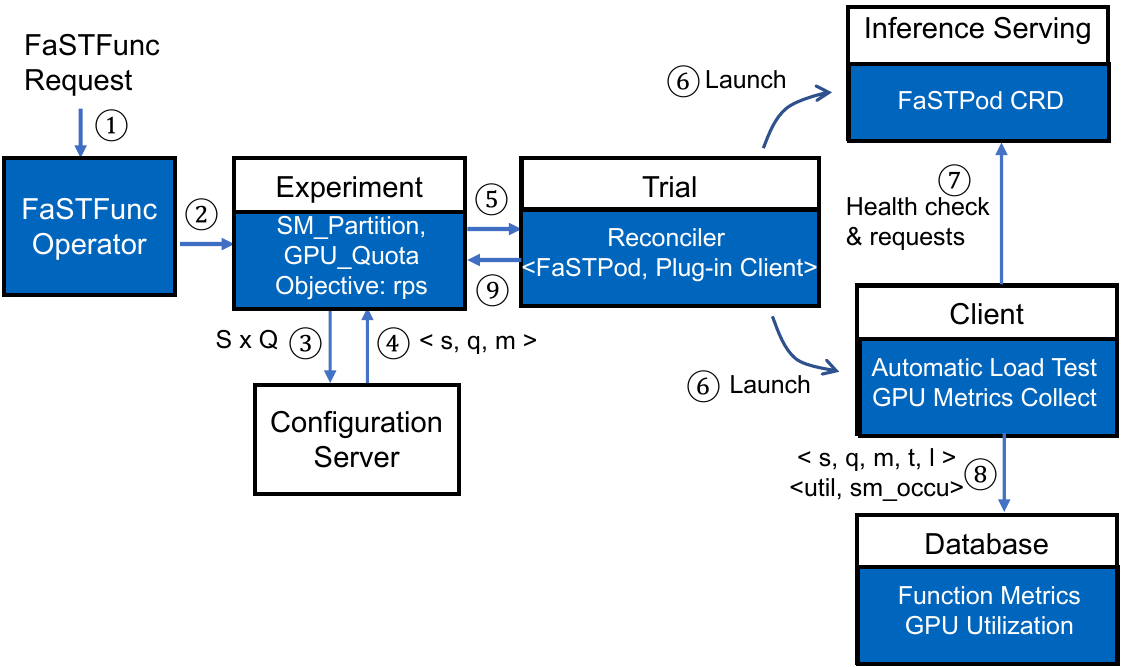}
    \caption{The workflow of the FaST-Profiler.}
    \label{fig:profiler}
  \end{minipage}
  \vspace{-2mm}
\end{figure*}}

The design and workflow of FaST-GShare is shown in Figure \ref{fig:architecture}. The architecture follows the Kubernetes cluster design for FaaS compatibility and consists of four key components: the FaST-Profiler and FaST-Scheduler in the control plane, and the FaST-Manager and Model Sharing Storage in work nodes. 

FaST-Profiler automates the process of profiling function throughput across various spatio-temporal resource allocations. When triggered by function images or built codes, it launches the FaSTPod to perform deep learning inference tasks, generates load testing requests, and collects metrics from Metric Collector. The profiling data is then stored in the database. Based on this profiling information, the FaST-Scheduler allocates efficient spatio-temporal resources to pods, controls auto-scaling of FaSTPods based on predicted request loads from the gateway, and schedule functions to efficient GPU nodes. FaSTPod is the customized controller that allocates spatio-temporal resources to inference pods of a function. The FaST Frontend in the pod access GPU resources by sending token requests and receiving time tokens from FaST-Manager. Model Storage is designed to share model tensors among pods of the same function to alleviate memory contention. More details about the design of each component are given in the following sections.

\subsection{FaST-Profiler}
As different functions have their own performance curves with regard to the resources, function profiling is needed for the subsequent scheduler to guarantee SLOs. Therefore, we propose FaST-Profiler, an automatic and flexible FaaS profiler, that profiles the function throughput under various spatio-temporal GPU resource allocations. FaST-Profiler follows the basic structure of the Morphling Configuration Framework \cite{morphing_socc_2021}.  But Morphling only aimed at CPU resources and lacks flexibility for automatic profiling of FaaS functions. To adapt the spatio-temporal GPU sharing mechanism, each module of the Morphling has been re-designed.

The architecture of FaST-Profiler follows an Experiment-Trial workflow, as illustrated in Figure \ref{fig:profiler}. The Experiment phase involves configuring resources and initiating a profiling Trial. Resource configurations are sampled from the Configuration Server. The Trial phase is responsible for launching both the Client to generate the test workload and the FaSTPod to perform inference tasks. During this phase, the Client collects various metrics, such as function throughput, latency, and GPU metrics like utilization and SM occupancy, and stores them into databases. In FaST-Profiler, each module is reconstructed based on the Custom Resource Definition (CRD) \cite{crd_introduction} to support spatio-temporal GPU resource allocation. The FaSTFunc CRD is designed to automatically specify and wrap the function image created from the user's code. The FaSTFunc Operator is used to initiate the Experiment phase. 

% \subsubsection{Automatic and Flexible Client}
% The client in the FaST-Profiler achieves automatic and flexible processes for load generation, profiling evaluation, and metrics collection. Unlike the original client, which required users to manually implement these steps and repeatedly construct the client, FaST-Profiler separates these functional components and multiplexes the client through dynamic configuration passing. Additionally, a Logic Control API is provided to encapsulate testing details and enable adaptive load generation, concurrent virtual users, load ramping pace and runtime definition, termination conditions, and automatic metrics collection. This approach allows the system to meet various profiling requirements of functions efficiently.

\begin{figure}[htbp]
\begin{myVerbatim}
apiVersion: faasshare.com/v1
kind: FaSTPod
metadata:
  annotations:
    faasshare/sm_partition: "12"
    faasshare/quota_limit: "0.8"
    faasshare/quota_request: "0.3"
    faasshare/gpu_mem: "1073741824"
  name: fastsvc-rnnt-q30-p12
spec:
  podSpec:
    containers:
    - env:
      - name: MODEL_NAME
        value: MLPerf-FaaS-rnnt
      image: xxxx/mlperf-faas-rnnt:latest
  replicas: 2
\end{myVerbatim}
\vspace{-1mm}
\caption{A FaSTPod CRD controller specification.}
\label{code:fast_pod_spec}
\vspace{-6mm}
\end{figure}

% \subsubsection{FaSTPod}
% FaSTPod is responsible for autoscaling the subsequent GPU pod.
\textbf{FaSTPod} is a customized CRD controller used for managing pods with spatio-temporal sharing and scaling in Kubernetes. It serves as the basic unit for function deployment in FaST-Profiler and FaST-Scheduler. Unlike the traditional Deployment controller that can only allocate an integer number of GPUs to a pod, FaSTPod supports the specification of arbitrary fractions and combinations of spatial and temporal GPU resources. For example, 12 \textit{sm\_partition} stands for 12\% SMs of a GPU, 0.8 \textit{quota\_limit} and 0.3 \textit{quota\_request} denote the maximum and minimum timing quotas (800ms and 300ms if the time window is \textit{1s}) assigned to each pod, and \textit{gpu\_mem} is the memory allocation. Moreover, different from the SharePod specification of previous work \cite{kubeshare_hpdc_2020}, these resource configurations are automatically filled by FaST-Profiler and FaST-Scheduler, eliminating the need for manual user input. When a FaSTPod needs to scale up/down a function pod, it automatically registers/releases allocated spatio-temporal GPU resources in FaST-scheduler and synchronizes the resource configuration to the FaST Backend table as described in section \ref{sec:fast_manager}.

\subsection{FaST: Spatio-Temporal GPU Sharing Manager}
\label{sec:fast_manager}
% \begin{figure}[h]
%   \centering
%   \includegraphics[width=0.90\linewidth]{acmart-FaST/samples/Architecture/pics/FaST_Manager_v2.png}
%   \caption{}
%   \Description{}
% \end{figure}
% \begin{figure*}[th]
%   \centering
%   \includegraphics[width=\linewidth]{acmart-FaST/samples/Architecture/pics/FaST_scheduler.png}
%   \caption{}
%   \Description{}
% \end{figure*}
\begin{figure*}[htbp]
  \centering
  \begin{subfigure}[b]{0.35\textwidth}
    \includegraphics[width=\linewidth]{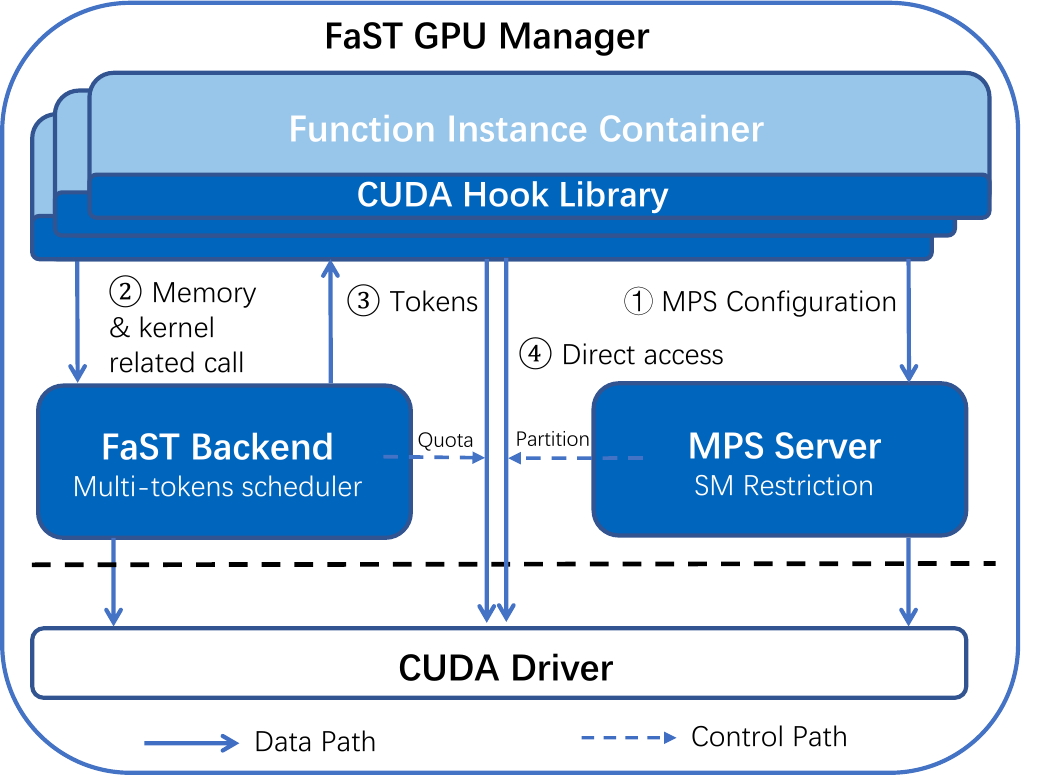}
    \caption{FaST GPU Manager.}
    \label{fig:fast_manager}
  \end{subfigure}
%   \quad % or any other spacing command you like
  \begin{subfigure}[b]{0.64\textwidth}
     \includegraphics[width=\linewidth]{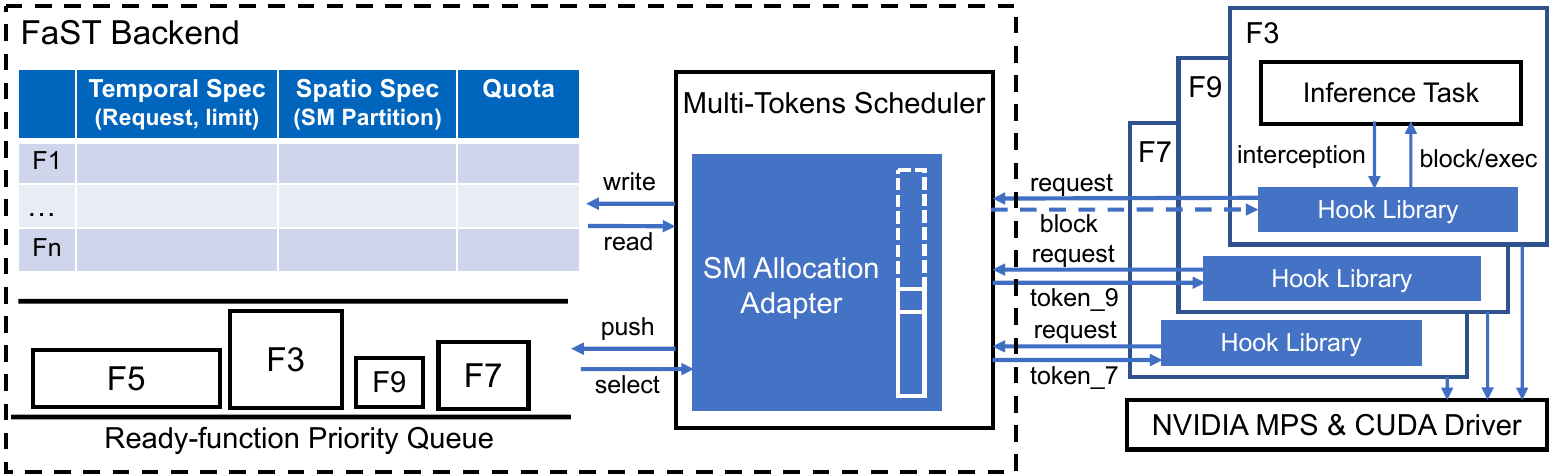}
    \caption{FaST Backend with Multi-tokens Scheduler.}
    \label{fig:fast_backend_scheduler}
  \end{subfigure}
  \vspace{-3mm}
  \caption{The mechanism of FaST-Manager to enable spatio-temporal GPU sharing cooperation.}
%   \label{fig:intro_gpu_utilization}
  \vspace{-3mm}
\end{figure*}

The spatio-temporal GPU sharing manager (FaST-Manager) is a mechanism designed to limit, prioritize, and isolate GPU resource usage in both spatial and temporal dimensions. The goal is to enable efficient sharing of GPUs among different pods. Unlike the \textit{cpuset} and \textit{cpu} subsystems in \textit{cgroups}, which have hardware and system kernel support to limit CPU usage, the GPU does not provide corresponding mechanisms, and the majority of NVIDIA GPU's underlying implementations are closed-source. To address this challenge, we construct a Frontend-Backend architecture based on existing NVIDIA toolkits to integrate and abstract spatial and temporal GPU multiplexing. As shown in Figure \ref{fig:fast_manager}, the frontend refers to the function instance container with the CUDA hook library. The container resides within the pod of FaSTPod and accommodates deep learning tasks. The backend involves FaST Backend with multi-tokens scheduler and MPS Server Backend, which manage temporal and spatial GPU resources respectively. 

Figure \ref{fig:fast_manager} illustrates the collaboration process between the frontend and backend. \textcircled{1} When the function instance container receives the spatio-temporal GPU configuration from FaSTPod, it initializes the corresponding SM partition in the MPS server. \textcircled{2} Meanwhile, the container registers the allocated time quota and memory in the FaST Backend. When the deep learning task starts executing, its calls to CUDA Driver APIs are intercepted by the CUDA Hook Library to request time tokens from the FaST-Backend. \textcircled{3} The FaST Backend evaluates the request and determines whether to dispatch the token. \textcircled{4} If the token is granted, the deep learning task can start accessing CUDA driver for kernel launching, with timing token and SM partition limitation in effect. To ensure flexible resource scheduling and necessary elasticity, we adopt the mechanism of controlling spatial multiplexing through timing tokens to achieve the spatio-temporal limitation collaboration in the FaST manager.

\subsubsection{MPS-based Spatial Sharing Backend}
Most data center GPUs possess lots of SMs and CUDA cores, but they are typically not efficiently utilized in deep learning inference. Multi-Process Service (MPS) enables multiple CUDA processes to share a GPU by the Hyper-Q mechanism, which allows CUDA kernels to be processed concurrently on the same GPU. Meanwhile, MPS provides performance isolation mechanisms to avoid process interference. Therefore, we leverage MPS to design the spatial sharing backend. The MPS exploits the server-client architecture, where the server provides shared connections and concurrency to the clients while the client is the runtime library providing interfaces for applications to communicate with the server. Thus, our backend aims to integrate the MPS components into our Manager to achieve spatial resource isolation and abstraction for FaSTPod.  

The spatial sharing backend constructs a container managed by the DaemonSet controller to accommodate the MPS server for each GPU node. The container exposes the IPC namespace for clients' connection and initializes the exclusive MPS mode. As for the client, it is integrated into the runtime of FaSTPod. Upon initialization of the function pod, FaSTPod automatically enables the corresponding IPC namespace and configures the SM partition for the client through an environment variable, thereby restricting the available spatial resources for the pod. MPS provides two partitioning strategies for SM configuration, namely Programming Interface and Environment Variable \cite{mps_introduction}. However, the programming interface requires injecting the driver API call $cuCtxCreate\_v3$ into ML frameworks while performing the configuration, which conflicts with our framework-agnostic design. Furthermore, some frameworks utilize different driver APIs to create CUDA context. Therefore, the environment variable, CUDA\_MPS\_ACTIVE\_THREAD\_PERCENTAGE, is exploited to isolate spatial resource usage for FaSTPod. %The strategy prioritizes distributing the work submitted by different clients to non-overlapping SM sets, effectively reducing interference among the work submitted by different clients.

\subsubsection{FaST Time Sharing Backend}
The FaST Backend is designed to perform and limit temporal GPU multiplexing among pods with the consideration of the spatial sharing. Following the widely used vGPU resource limitation architecture in previous works \cite{gu2018gaiagpu, chen2021gemini, kubeshare_hpdc_2020}, we adopt the monitoring-scheduling mechanism, where the monitor collects real-time GPU usage data for each pod, while the scheduler determines whether to permit further execution based on each pod's usage and the corresponding time quota specified in FaSTPod.

% the monitor collects the amount of GPU temporal usages of pods in real-time, and scheduler determines whether to allow pods' further execution on the GPU based on their usages to meet corresponding temporal resource allocations, namely the time quota specified in FaSTPod. 

As shown in the Figure \ref{fig:fast_backend_scheduler}, the mechanism requires the cooperation between the hook library in the inference pod and the FaST Backend. The hook library intercepts the CUDA driver APIs, such as \textit{cuLaunchKernel} and \textit{cuCtxSynchronize}, using the Linux LD\_PRELOAD mechanism. It sends the token request to the FaST backend asking for further CUDA kernel executions if the dispatched time token of the pod has been consumed. The token represents the time slice available to launch CUDA kernels during deep learning inference. Whenever one or more CUDA kernels complete executions on the GPU, CUDA needs to perform a host-side synchronization by calling a CUDA synchronization API, such as \textit{cuCtxSynchronize} and \textit{cuMemcpyDtoH}. To monitor resource usage with high granularity, a CUDA time counting event is inserted before calling the synchronization API, allowing flexible and fine-grained timing. This event monitoring and token mechanism was first proposed in Gemini \cite{chen2021gemini} and used in Kubeshare \cite{kubeshare_hpdc_2020}. However, they only support the passing of a single token among pods for GPU timing sharing, making it impossible to schedule multiple concurrent pods with spatial resource multiplexing in our FaST manager. Thus, we integrated the mechanism to design the FaST Backend and enable multi-token scheduling.

The Multi-tokens Scheduler is capable of receiving and dispatching multiple tokens simultaneously for pods that are running concurrently with spatial partitions. In the FaST Backend, the table stores the time quota ($Q_{used}$) used for each function pod and also stores the temporal and spatial resource configurations ($Q_{request}$, $Q_{limit}$, and $S_{SMs}$) from either the FaST-Scheduler or the FaST-Profiler. When function pods send token requests to the scheduler, the Multi-Token Scheduler performs three operations: filtering, candidate pod enqueuing, and token dispatching. Filtering involves calculating the remaining time quotas of each pod, namely $Q_{miss} = Q_{request} - Q_{used}$ and $Q_{remain} = Q_{limit} - Q_{used}$. Pods with $Q_{remain} \leq 0$ are blocked and must wait for the next time window due to exceeding the maximum allocated time quota (such as $F_3$), while other pods are enqueued into the Ready-function Priority Queue. This queue sorts pods in descending order by $Q_{miss}$, ensuring that the scheduler always prioritizes scheduling pods with the largest timing missing gap. Since there is a spatial resource limitation for each pod and over-allocation of SMs can cause performance interference, the SM Allocation Adapter ensures that the current system's SM occupancy does not exceed the $SM\_GLOBAL\_LIMIT$ limit of $100\%$. Therefore, the adapter continuously returns tokens for the head pods in the queue until it encounters $S_{SMs} + S_{running} > 100\%$. Multiple pods with different SM partitions can execute concurrently on the GPU. Hence, time multiplexing controls the spatial resource sharing.

The variables \textit{quota\_limit} ($Q_{limit}$) and \textit{quota\_request} ($Q_{request}$) specify the maximum and minimum GPU temporal resource allocation for a pod, respectively. Our architecture implements elastic resource allocation, which is similar to Kubernetes and previous research \cite{kubeshare_hpdc_2020}, to enable pods to utilize more GPU resources when the GPU is idle, thus improving overall GPU utilization. For profiling purposes in the FaST-Profiler, both values are set to the same value.

\vspace{-7pt}
\subsection{FaST-Scheduler}
Different from traditional scenarios that users define pods' resource allocation for the scheduler, FaaS requires the scheduler to automatically allocate resources, and then scale and schedule functions across nodes to guarantee SLOs.  In this section, we introduce FaST-Scheduler that can efficiently utilize GPU under the spatio-temporal sharing mechanism while ensuring function SLOs. The process of the scheduler is mainly divided to two parts, namely the heuristic auto-scaling and the node selection. The heuristic auto-scaling leverages the profiling information from the FaST-Profiler to heuristically decide the number of function pods and corresponding efficient resource allocations under various workloads. The node selection phase aims to effectively schedule function pods to GPUs while performing the 2D resource restructuring and planning of GPUs to enhance GPU resource efficiency and mitigate resource fragmentation.

\begin{algorithm}[hbt]
\SetAlgoLined
\caption{The Heuristic Scaling Algorithm}
\small
\label{alg:heuristic_schedule}
\KwIn{ $\Delta RPS$, the RPS processing gap of all functions; Profile $P$ = $\{<F_j, S_{j,p}, Q_{j,p}, T_{j,p}>\}$; Priority queues $\{L_j\}$ for each running function $F_j$, pods in $L_j$ are sorted in ascending order by $RPR$.}
\KwOut{cfgs: new function configurations list; }
cfgs $\gets []$ \;
\For{$F_j, \Delta RPS_j$ \textbf{in} $\Delta RPS$}{
    \eIf{$ \Delta RPS_j \geq 0 $}{
        $ p_{eff} \gets \arg\max_{p} \frac{T_{j,p}}{S_{j,p} \, * \, Q_{j,p}} $ \;
        $T_{eff} \gets T_{p_{eff}}$ \;
        $\textbf{let} \,\, \Delta RPS_j = n \cdot T_{eff} + r_j$  \textbf{where} $n=\lfloor \frac{\Delta RPS_j}{T_{eff}} \rfloor, r_j >0$ \;
        cfgs $\gets$ cfgs \, $\bigcup$ \, $n  \times <F_j, S_{j, p_{eff}}, Q_{j, p_{eff}}, +>$ \; 
        $L_j$ \textbf{pushes} $n$ times of $<F_j, S_{j, p_{eff}}, Q_{j, p_{eff}}, T_{j,p_{eff}}>$ \;
        $p_{ideal} \gets \arg\min_{p} (T_{j,p} - r_j)$ \textbf{where} $T_{j,p} > r_j$ \;
        cfgs $\gets$ cfgs \, $\bigcup$ \, $<F_j, S_{j, p_{ideal}}, Q_{j, p_{ideal}}, +>$ \;
        $L_j$.\textbf{push}( $<F_j, S_{j, p_{ideal}}, Q_{j, p_{ideal}}, T_{j,p_{ideal}}>$) \;
    }{
      $ \Delta R  \, \gets \, \Delta RPS_j $ \;
      \While{$\Delta R$ < 0}{
        $ <J_i, S_{J_i}, Q_{J_i}, T_{J_i}> \, \gets \,$ $L_j$.\textbf{front()} \;
        \If{$ \Delta R + T_{J_i} \leq 0 $}{
            cfgs $\gets$ cfgs \, $\bigcup$ \, $<J_i, S_{J_i}, Q_{J_i}, ->$ \;
            $L_j$.\textbf{pop()} \;
            $\Delta R \, \gets \, \Delta R + T_{J_i} $ \;
        }
      }
    }
    \Return cfgs \;
    \tcp{"<+>": scaling-up;  "<->": scaling-down;}
}
\end{algorithm}

% \begin{algorithm}
% \caption{Addition}\label{addition}
% \begin{algorithmic}[1]
% \Procedure{Add}{$a,b$}\Comment{ \\ \textbf{Input:} two numbers $a$ and $b$}
% \State $c \gets a + b$
% \State \textbf{Output:} sum of $a$ and $b$ is $c$
% \State \textbf{return} $c$
% \EndProcedure
% \end{algorithmic}
% \end{algorithm}

\subsubsection{Heuristic Auto-Scaling} The Auto-scaling module scales functions dynamically to adapt to traffic changes to meet SLOs. Owing to FaST-Profiler, we can determine the request processing capabilities ($T_{j,i}$) of each function ($F_j$) based on the resource allocation of SM partitions $S_{j,i}$ and time quotas $Q_{j,i}$ for each running pod $J_i$. Further, the processing gap of a function $F_j$ in RPS (Requests Per Second) can be quantified under different predicted request loads ($R_j$).
\vspace{-2mm}
$$
\label{formu:delta_rps}
\Delta RPS_j = R_j - (\sum_{J_i \in F_j} T_{j,i}) 
\vspace{-1mm}
$$
The gap value $\Delta RPS_j$ might be either positive or negative, where positive values indicate the overload that needs functions' scaling-up to feed these requests while negative values represent the over-assignment that requires the scaling-down to avoid ineffective resource occupation. Then how to select effective configurations for functions to scale becomes the key problem.  As shown in Algorithm \ref{alg:heuristic_schedule}, we define the metric $RPR$ (RPS per Resource) $ = \frac{T_{j,p}}{S_{j,p} \, * \, Q_{j,p}} $ to evaluate the GPU processing efficiency of different spatio-temporal resource combinations in a function, where $T_{j,p}$ denotes the throughput of the function $F_j$ with the configuration $p$. During functions' scaling-up, the scheduler always prioritizes choosing the most efficient resource combination $p_{eff}$ with $n$ pods to handle the most majority of requests and selects the minimum but sufficient resources $p_{ideal}$ for residual requests. Moreover, because $RPR_{p_{ideal}}$ is always lower than $RPR_{p_{eff}}$, and priority queue $L_j$ sorts function instances following the ascending order by $RPR$. In scaling-down scenarios, the scheduler prioritizes maintaining the most efficient function instances and delete residual-request instances with lower corresponding throughput efficiency. Hence, efficient function pods have a greater chance and a longer duration to handle requests, enabling efficient fine-grained GPU usages at the scheduler level.

% In the FaST-Profiler, different GPU SM partition and time quota allocations for a function results in different throughput.     
% \subsubsection{Heuristic Auto-Scaling}The Auto-scaling module employs a dynamic scaling strategy to effectively handle traffic fluctuations and satisfy function SLOs. Leveraging the FaST-Profiler, we can ascertain the total request processing capabilities ($\sum_{J_i \in F_j} T_{j,i}$) of each active function instance $F_j$ denoted by $T_{j,i}$, which is determined based on the allocation of resources to SM partitions $S_{j,i}$ and the time quota $Q_{j,i}$ assigned to each pod $J_i$. This approach enables us to quantify the Requests Per Second (RPS) processing gap of a function $F_j$ under varying request loads $R_j"$.

\subsubsection{Node Selection}

\begin{algorithm}[hbt]
\SetAlgoLined
\caption{The Maximal Rectangles Algorithm}
\small
\label{alg:max_rectangle}
\KwIn{$G = n \times (W,H)$, $n$ GPUs; \, $F = (w,h)$, a function pod waiting for scheduling; \, $L =\{L_1, L_2, L_i, ...\}$, the free rectangle lists, where  $L_i =\{R_i^{j}\}$, free rectangles in $G_i$; }
\KwOut{$L^{*}$,  updated free rectangle list of $G_t$; \, $<G_t, F>$, node binding between $G_t$ and $F$.}

$R_t^{k} \gets \arg\min_{R_i^{j}} (Area(R_i^{j}) - Area(F))$,  
\, s.t. $w_{R_i^{j}} \leq w_F$ and \qquad  \qquad  \qquad $h_{R_i^{j}} \leq h_F $; \\*
\If{$R_t^{k}$ \textbf{not} exists}{
    \Return A new GPU required;
}
$R_{t}^{'}, R_{t}^{''} \, \gets \, PlaceAndNewJointRect(G_t, \, R_t^{k},F, "BottomLeft")$ \;
$L_{t} \, \gets \, (L_t - R_t^{k}) \bigcup \, \{R_t^{'}, R_t^{''}\}$ \;

\tcp{Intersection update;} 
$L^{*} \, \gets \, L_t$ \;
\For{\, $R \, \, \textbf{in} \, \, L_{t} $}{
    $I \, \gets \, R \bigcap F$ \;
    \If{$\, I \, \neq \, \emptyset$}{
        $\{R^{1}, R^{2}, R^{3}, R^{4}\} \, \gets \, Subdivide(R, I)$ \;   
        $L^{*} \, \gets \, L^{*} \bigcup \{R^{1}, R^{2}, R^{3}, R^{4}\}$ \;
    }
}
\tcp{Remove redundant free rectangles;} 
\For{ \textbf{each} pair $(R^{j}, R^{p})$, \textbf{where} \, $R^{j}, R^{p} \in L^{*}$}{
    \If{$R^{j} \subseteq R^{p}$}{
        $L^{*} \, \gets \, L^{*} - R^{j}$ \;
    }
}
\Return $L^{*}$, $<G_t, F>$ \;
\end{algorithm}

Once FaSTPod configurations are received, node selection focuses on scheduling pods to GPU nodes with 2D resource allocation to efficiently utilize GPU computing resources. However, fine-grained resource allocation can easily lead to resource fragmentation. To mitigate this issue, Kubernetes and previous work \cite{kubeshare_hpdc_2020} have introduced locality constraints, such as affinity and exclusion. Nevertheless, as the partitioning dimension increases, resource fragmentation becomes worse and poses greater challenges due to larger search spaces. To address these challenges, inspired by the 2D Bin packing algorithm \cite{jylanki2010thousand}, we propose the Maximal Rectangles Algorithm (MRA), a novel mechanism that dynamically plans and reorganizes GPU resources and binds pods to GPUs to achieve efficient GPU scheduling with low fragmentation.

\begin{figure}[ht]
  \centering
  \includegraphics[width=0.80\linewidth]{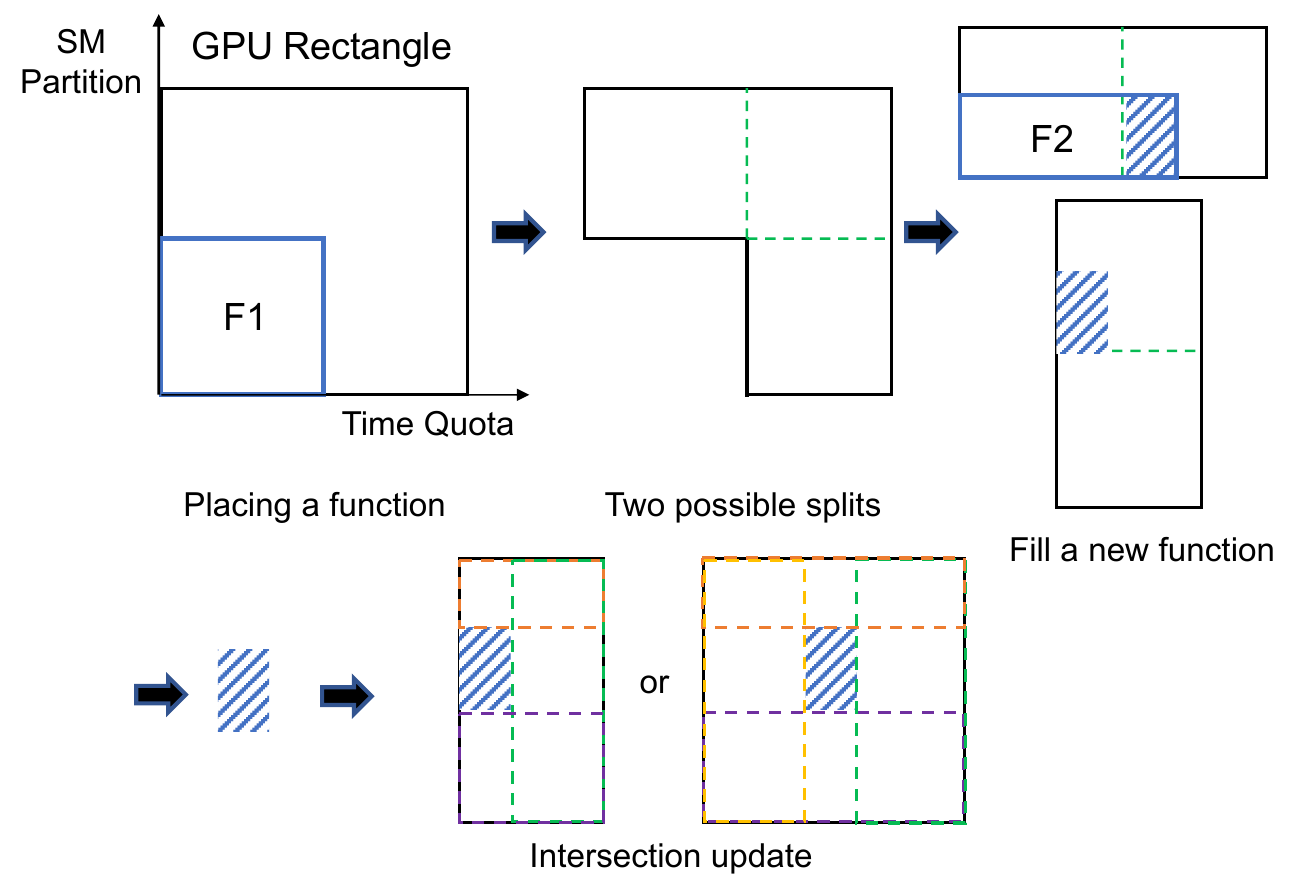}
  \vspace{-3mm}
  \caption{$PlaceAndNewJointRect$ and $Subdivide$ operation.}
  \label{fig:max_rectangle_diagram}
  \Description{}
  \vspace{-2mm}
\end{figure}

As shown in Figure \ref{fig:max_rectangle_diagram}, the time and spatial dimensions of the GPU resource are formalized as the width and height of a rectangle, i.e., $W*H = 100\% \,quotas * 100\% \, SMs$ for a GPU. Each GPU ($G_i$) has a list ($L_i$) to store free rectangles which represent the available resources for new pods. Node selection involves two steps: rectangle best matching for node selection and resource rectangle restructuring. First, as described in Algorithm \ref{alg:max_rectangle}, for a new function pod, the scheduler searches all rectangle lists on the GPUs to identify a rectangle that satisfies the pod's temporal and spatial requirement while having the minimum difference in "secondCores" value (line 1). The "secondCores", denoted as the \textit{Area}, is defined to uniformly evaluate the size of GPU spatio-temporal resources, i.e. $Quota \, \times \, SMs$. If the optimal rectangle ($R_t^{k}$) exists, the pod rectangle is scheduled to the corresponding GPU $G_t$. The global best matching mechanism is able to prioritize scheduling pods to the GPUs that already have resource rectangles rather than to new GPUs and allow the GPU shared in time and spatial dimension to be maximally occupied by pods, which significantly improves GPU efficiency.

The resource rectangle restructuring can be divided to three steps, including split, intersection update, and redundant rectangle removing. As depicted in Figure \ref{fig:max_rectangle_diagram}, after accommodating the pod rectangle $F_1$, the original resource rectangle has two possible splits. To maximize the resource usage, we always keep the two maximum rectangles rather than a specific type of divisions in the rectangle list. The "maximal" in the MRA refers to the fact that the new rectangles $R_t^{'}$ and $R_t^{''}$ are the largest possible rectangles in each direction. However, since the free rectangles in the list are not mutually exclusive, the scheduler performs the intersection update for all rectangles to remove the area allocated to the new function pod ($F2$). Subdividing the intersection areas creates three to four new resource rectangles, some of which may be redundant. To reduce search complexity and the possibility of fragmentation, smaller resource rectangles inside larger rectangles are merged.

Regarding the reclamation of resource rectangles, we implement a "keep-restructure" policy, whereby released rectangles from pods are added back to the list of free rectangles. This policy facilitates the seamless reuse of same resources for the same functions and reduces the likelihood of resource fragmentation. Once the number of rectangles in the list exceeds a certain threshold, the GPU is initialized as a single rectangle with dimensions of $W \times H$, and the rectangle restructuring above is launched to reorganize resource rectangles for pods running on the GPU, which prevents further resource fragmentation.

\vspace{-5pt}
\subsection{Model Sharing}
\vspace{-6mm}
\begin{figure}[htbp]
  \centering
  \includegraphics[width=0.47\textwidth]{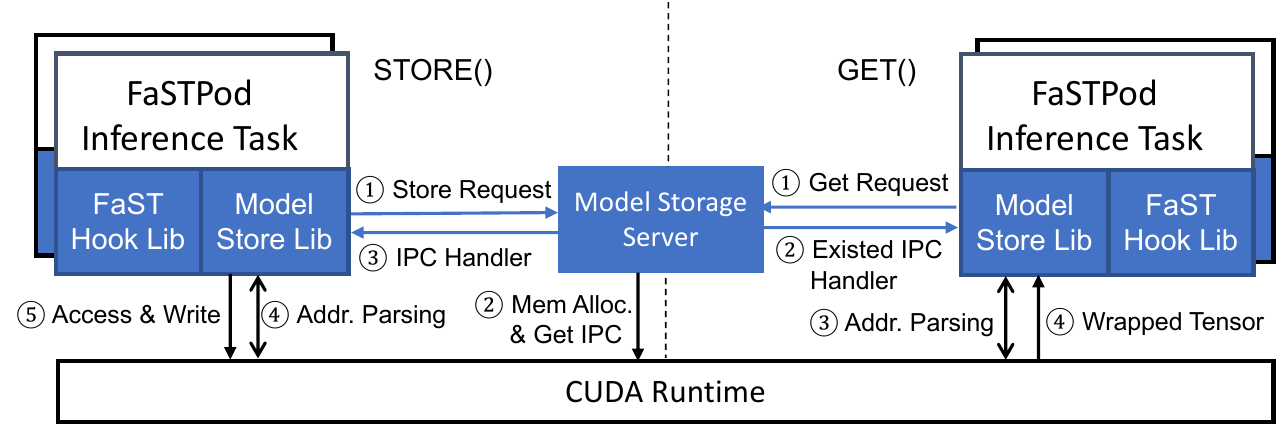}
  \vspace{-2mm}
  \caption{Model Sharing Mechanism.}
  \label{fig:model_sharing}
  \vspace{-3mm}
\end{figure}
The fine-grained GPU sharing mechanism allows more functions to use the same GPU, which can lead to memory contention among functions. To mitigate the problem, we propose the IPC-based (Inter-Process Communication) model sharing mechanism as illustrated in Figure \ref{fig:model_sharing}. The mechanism consists of two parts, namely the Model Store Lib and Model Storage Server. The model store library is designed to share model tensors between instances of the same function, so that an inference framework like PyTorch can construct the model with zero-copy. The Model Storage Server allocates the memory on the GPU and export the corresponding IPC handler to the model store library, so that the inference function can access it directly. As demonstrated in Figure \ref{fig:model_sharing}, we abstract the model sharing into two APIs, namely \textbf{\textit{STORE()}} and \textbf{\textit{GET()}}. In the \textbf{\textit{STORE}} phase, \textcircled{1} when a new function $F_A$ requires to allocate a tensor $T$ to the GPU memory, the Store Lib calculates the expected buffer size ($M_T$) of $T$ and then sends a store request with the buffer size and tensorID to the server. \textcircled{2} The storage server calls the CUDA driver API \textit{cuMemAlloc} to allocate GPU memory of size $M_T$ and \textit{cuIpcGetMemHandle} to acquire the IPC handle of the corresponding device memory pointer (devptr). Then \textcircled{3} the storage server transmits the IPC handle back to $F_A$. \textcircled{4} The Store Lib parses the IPC handle to get the devptr by driver API \textit{cuIpcOpenMemHandle} and finally \textcircled{5} write the data of Tensor $T$ to the devptr. In the \textbf{\textit{GET}} phase, \textcircled{1} when another instance $F_A^{1}$ of the $F_A$ tries to get the tensor $T$ with tensorID by sending a request to the storage server, \textcircled{2} the storage server checks the existence of $T$, and if $T$ exists, corresponding IPC handle is returned, otherwise, the \textit{STORE} operations are triggered. Then \textcircled{3} the Store Lib follows the same operations to parse the IPC handle, and \textcircled{4} the raw pointer is wrapped in a tensor object for the function instance $F_A^{1}$.  In this way, only one copy of the model tensors will reside on the GPU memory. Multiple function instances can share the model directly in a memory-efficient way.

\vspace{-5pt}
\section{Implementation Details}
We implemented FaST-GShare by extending modules in the OpenFaaS faas-netes \cite{faas-netes} serverless framework. In FaST-Profiler, Morphling framework \cite{morphing_socc_2021} was utilized as the fundamental structure to implement the automatic profiling modules, and the load testing tool Locust \cite{locust} was integrated to implement the client's Logic Control API. We followed the Kubebuilder \cite{kubebuilder} framework to implement custom resource controllers for each CRD, such as FaSTPod and the profilers' CRD modules. In FaST-Manager, we implemented a multi-token scheduler with domain sockets, similar to Kubeshare \cite{kubeshare_hpdc_2020} which used socket for single token communication.  In Model Sharing, we leveraged the Apache Plasma Object Store API  \cite{plasma_object_store} to implement the mechanism for the Model Storage Server. Additionally, we constructed CUDA tensor objects by implementing a PyTorch C++ extension based on the \textit{libtorch} C++ API.

\vspace{-5pt}
\section{Experimental Evaluation}
\subsection{Experiment Setup}
The experimental testbed for FaST-GShare utilizes $N1$ series VM instance nodes on the Google Cloud Platform. The master node features an Intel(R) Xeon(R) CPU @ 2.00GHz with 48 cores and 60GB RAM. 4 work nodes are deployed, each equipped with a 24-core CPU and an NVIDIA Tesla V100 GPU with 80 SM units, 640 Tensor cores, and 16GB device memory. All nodes run Ubuntu 20.04.5 LTS with Kubernetes v1.24.2 and OpenFaaS with faas-netes v0.14.2 and gateway v0.21.3. 

We exploited various deep learning applications from the standard MLPerf benchmark \cite{mlperf_benkmarks} provided by NVIDIA to create FaaS function benchmarks. These DL applications included ResNet for image classification, BERT for natural language processing, RNNT for speech recognition, and GNMT for neural machine translation. In addition, we introduced larger transformer models, including ResNeXt \cite{resnext_model} (vision) and ViT\_hug \cite{dosovitskiy2020image}, to evaluate the effectiveness of the model sharing module. The DL models were performed on two popular deep learning frameworks, PyTorch and TensorFlow. To simulate user requests, we used Grafana k6 \cite{k6_load_generator} as the load generator. The NVIDIA DCGM-Exporter and nvidia-smi were 
used to collect GPU utilization, SM occupancy, and memory usages.

\subsection{Profiling and Isolation}
% FaST-Profiler is able to automatically profile function throughput under various spatial and temporal resource configurations. 
The profiling points of resources in FaST-Profiler are as follows:
\begin{itemize}
\item Temporal dimension: 20\%, 40\%, 60\%, 80\%, 100\%.
\item Spatial dimension: 6\%, 12\%, 24\%, 50\%, 60\%, 80\%, 100\%.
\end{itemize}
In fact, any resource granularity can be profiled with FaST-Manager. Practices have shown that the throughput over temporal dimension is basically proportional, thus, equal intervals are chosen for evaluation. Considering models with small sizes, the SM partition starts from a minimum of 6\%, and smaller intervals are used. 

\begin{figure}[htbp]
\centering
\centering
\includegraphics[width=\linewidth]{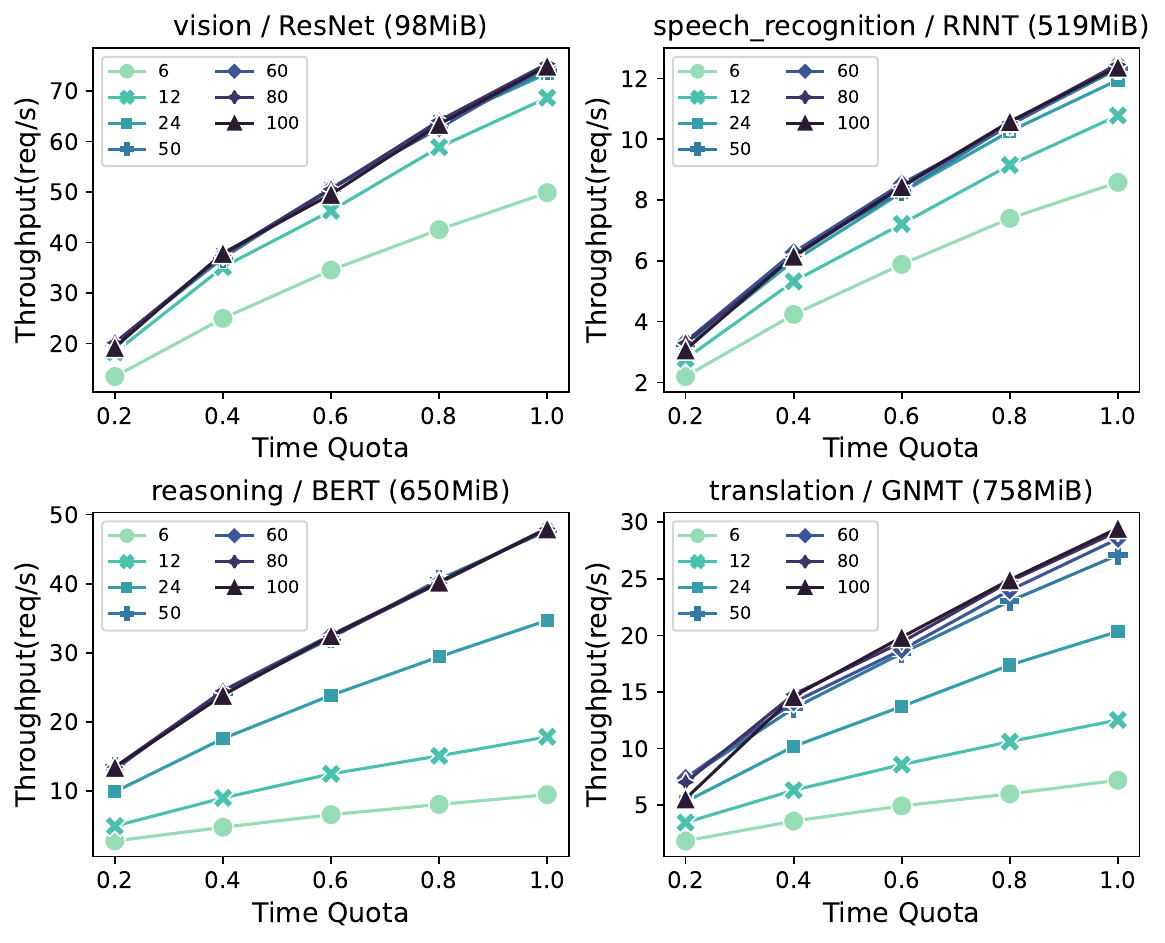}
\vspace{-6mm}
\caption{Function throughput of models from FaST-Profiler.}
\label{fig:profiling_result}
\vspace{-2mm}
\end{figure}

The profiling results are shown in Figure \ref{fig:profiling_result}. Observing from a temporal dimension, as the time quota increases, the throughput of all models increase proportionally. The limited throughput indicates that effective temporal resource isolation and limitation have been achieved. From a spatial dimension analysis, as the allocation proportion of SM partitions increases,  the models' throughput does not exhibit proportional growth. When the allocation of SM partitions increases to a certain point, the throughput will no longer increase. For example, in the ResNet model, an increase in SM partitions from 6\% to 12\% results in a much higher throughput improvement compared to an increase from 12\% to 24\%. When SM partitions reach 24\%, allocating more SM partitions does not result in a throughput increase. It indicates that a model cannot fully occupy all SMs, highlighting the necessity of spatial sharing in FaST-GShare. Furthermore, the results demonstrate that FaST-GShare is capable of effectively isolating and limiting spatial resources, while also enabling temporal control over the usage of these resources. Meanwhile, it can be observed that larger models require more SM partitions to reach the saturation state of spatial resources.

\subsection{Spatial Sharing Performance Validation}
The experiment is designed to evaluate the performance of spatial sharing with 100\% time allocation under different spatial sharing configuration: no spatial sharing (racing), 12\% partition, and 24\% partition. The over-subscription is allowed here to fully observe the characteristics of spatial sharing.

\begin{figure}[htbp]
\centering
\centering
\includegraphics[width=\linewidth]{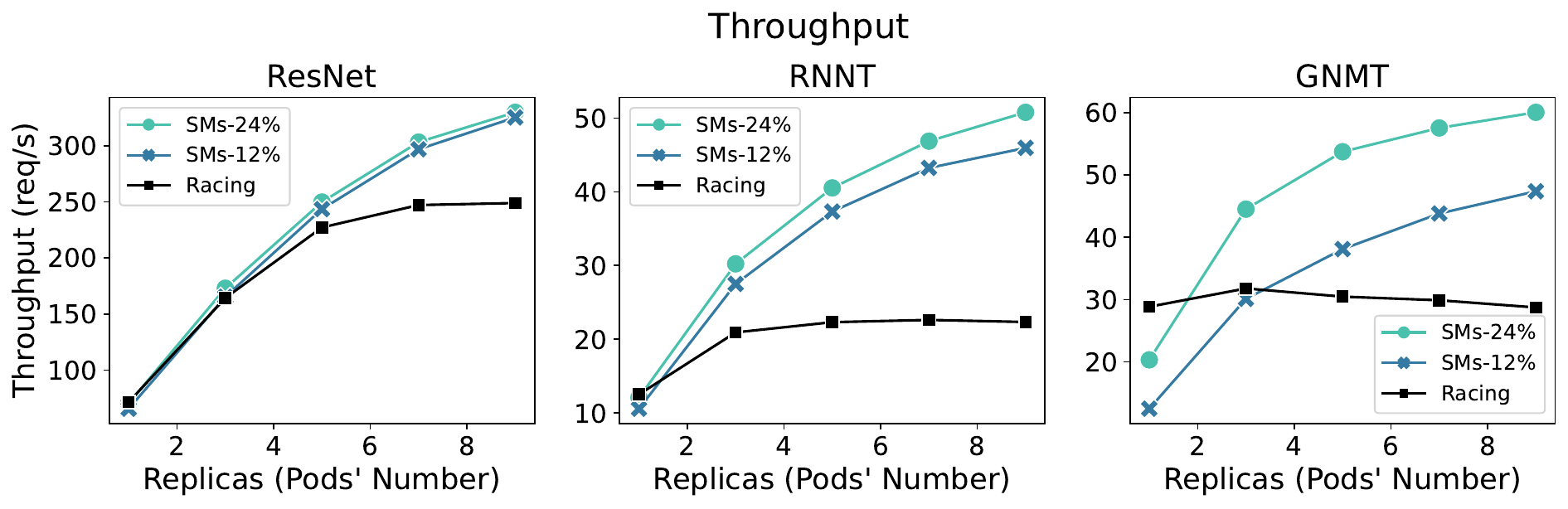}
\includegraphics[width=\linewidth]{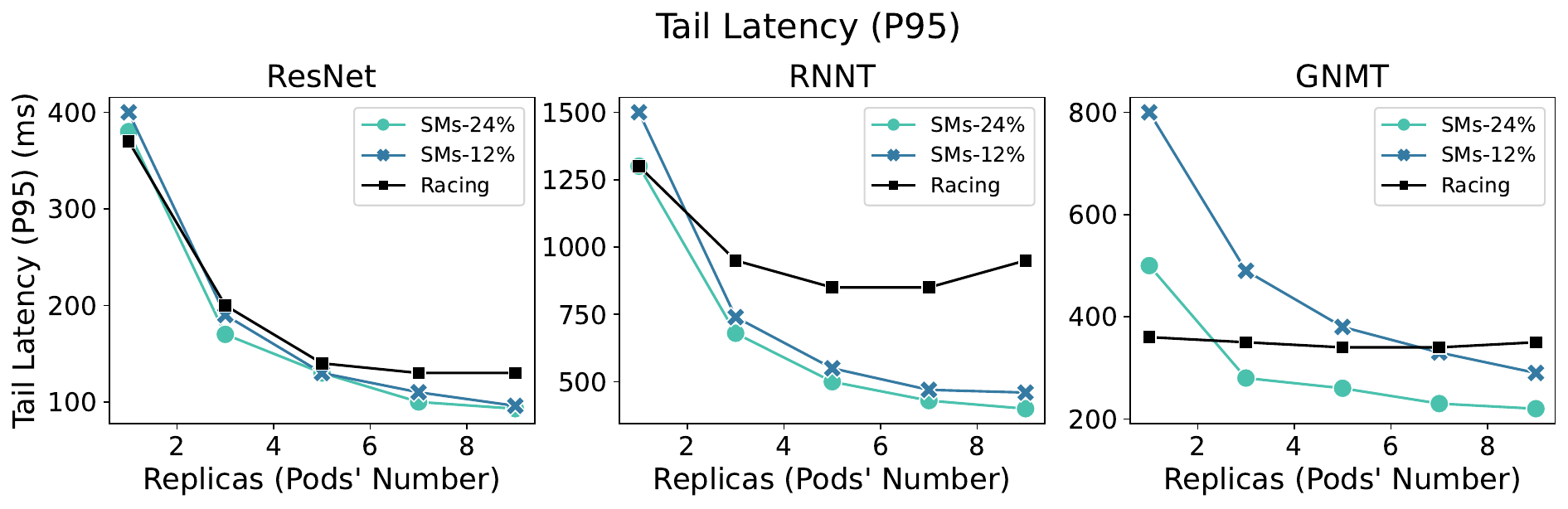}
\includegraphics[width=\linewidth]{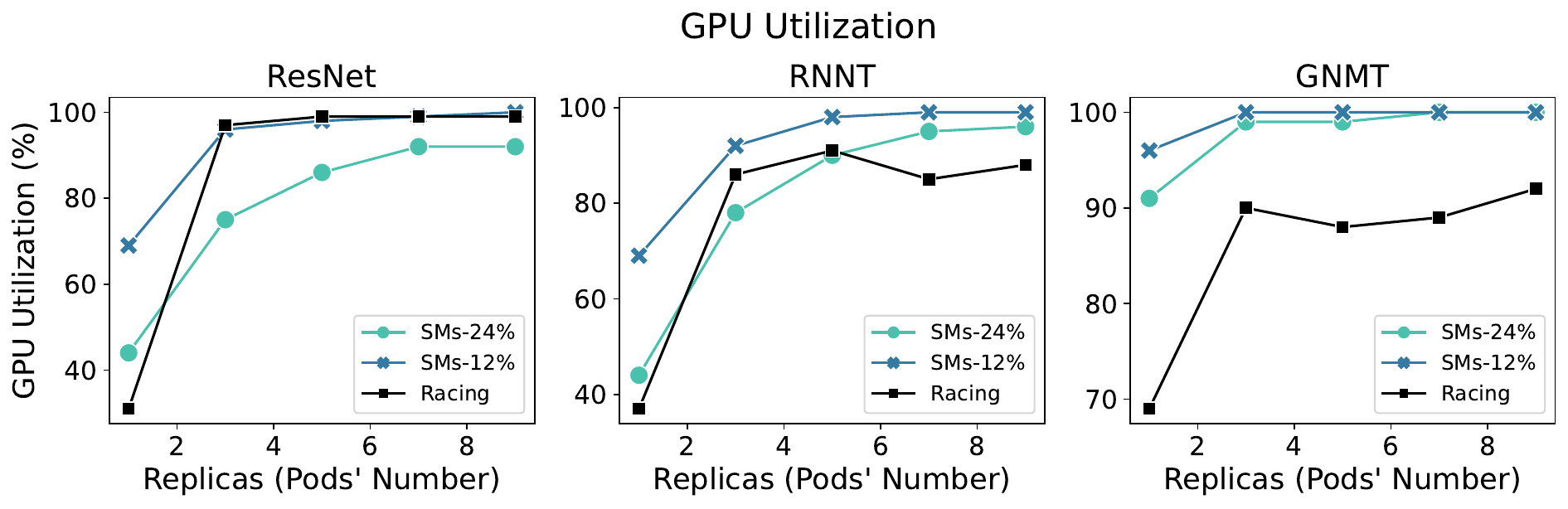}
\includegraphics[width=\linewidth]{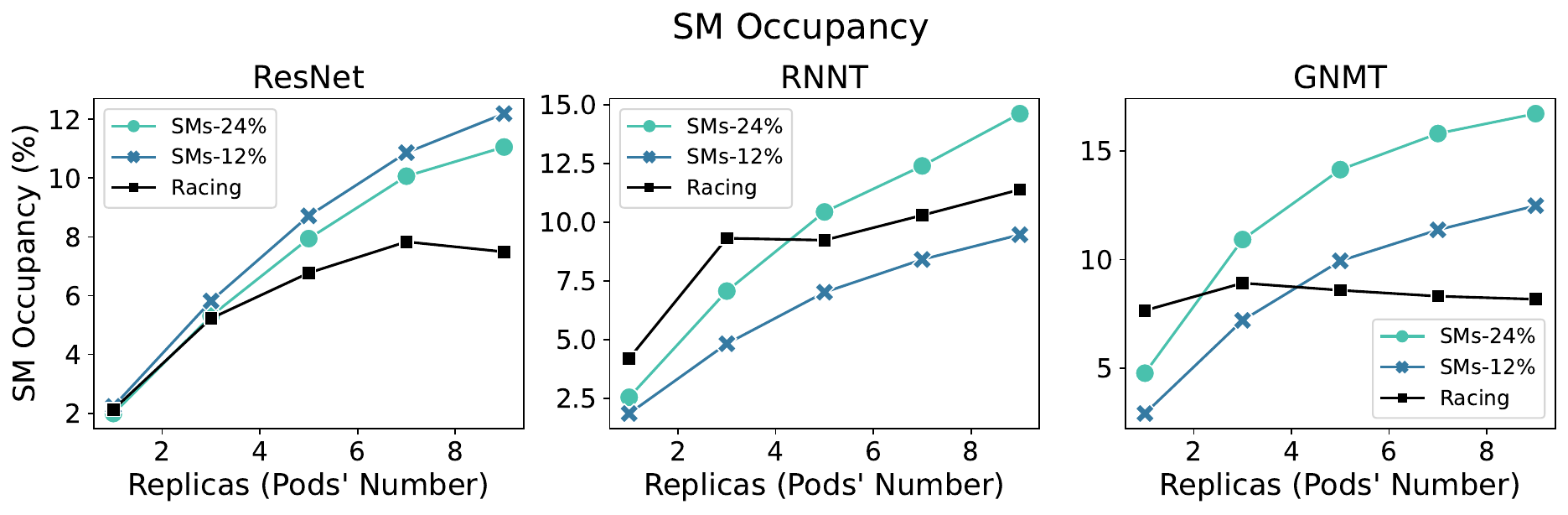}
\vspace{-6mm}
\caption{Performance of spatial sharing.}%名字怎么取是个问题
\label{fig:spatial_sharing_performance_result}
\vspace{-4mm}
\end{figure}

As shown in Figure \ref{fig:spatial_sharing_performance_result}, as the number of pods increases, spatial sharing of 12\% and 24\% partitions results in significant improvements in both throughput and SM occupancy, while also reducing tail latency accordingly. When the number of pods reaches a certain level, spatial sharing of the same number of pods can achieve much higher throughput, SM occupancy and utilization than racing along with much lower tail latency. 

\begin{figure}[htbp]
  \centering
  \begin{subfigure}[b]{0.22\textwidth}
    \includegraphics[width=\textwidth]{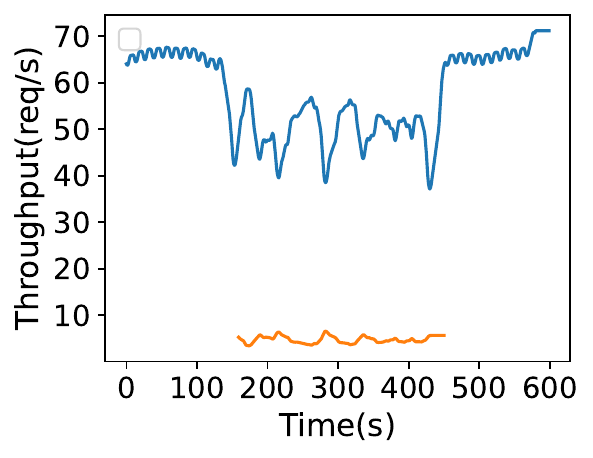}
    \vspace{-4mm}
    \caption{Only time sharing.}
    \label{fig:time_sharing_variable_control}
  \end{subfigure}
%   \quad % or any other spacing command you like
  \begin{subfigure}[b]{0.22\textwidth}
    \includegraphics[width=\textwidth]{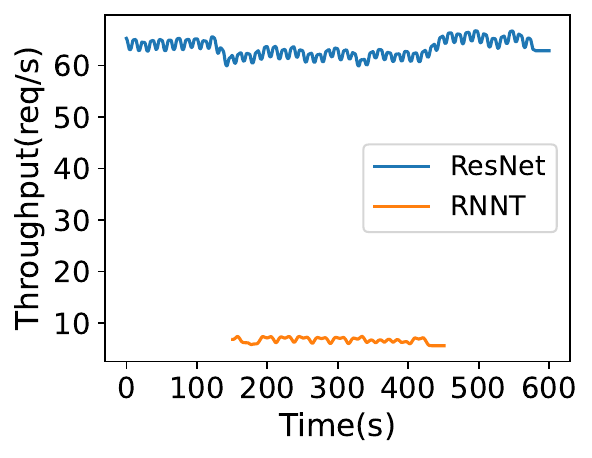}
    \vspace{-4mm}
    \caption{Spatio-Temporal sharing.}
    \label{fig:space_sharing_variable_control}
  \end{subfigure}
  \vspace{-2mm}
  \caption{Effectiveness of spatial sharing. }
  \label{fig:control_variable_test}
  \vspace{-2mm}
\end{figure}

Considering no over-subscription to prevent function interference, a single GPU can deploy only one racing pod, while it can deploy up to eight pods with a 12\% SM partition. Experimental findings suggest that spatial sharing yields significant performance advantages in all aspects. For instance, when using RNNT, eight pods with spatial sharing can achieve a throughput of 40 req/s, tail latency below 500ms, utilization up to 98\%, and SM occupancy nearly 10\%. Conversely, a single racing pod can only achieve a throughput of 12 req/s, tail latency above 1250ms, utilization below 40\%, and SM occupancy below 5\%, which falls significantly below spatial sharing. The maximum throughput achievable through time sharing is indicated by the throughput in a single racing pod. Therefore, the spatial sharing can achieve at least 3.15x (296.8 req/s and 71.37 req/s), 2.45x (43.24 req/s and 12.51 req/s), 0.52x (43.79 req/s and 28.85 req/s) higher throughput than time sharing for ResNet, RNNT and GNMT, respectively. The results indicate that spatial sharing in FaST-GShare can significantly improve function throughput, SM occupancy, and utilization while reducing tail latency.

We conducted a comparative experiment to validate the effectiveness of spatial sharing by comparing the performance impact of time sharing alone versus spatio-temporal sharing. ResNet and RNNT were configured with full SM quota but different time quotas (request-limit) in timing sharing (ResNet: 50\%-80\%, RNNT: 50\%-50\%). In the case of only time sharing, RNNT pod could interfere with the performance of ResNet pod due to elastic time quota (80\%+50\%>100\%). However, when spatial sharing was enabled and both were allocated the same SM partitions (24\%), there was no mutual influence between them. As shown in Figure \ref{fig:control_variable_test}, the experiment proved the effectiveness of spatial sharing in FaST-Manager.

\subsection{Efficient FaST-Scheduler}

\begin{figure}[htbp]
  \centering
  \begin{subfigure}[b]{0.23\textwidth}
    \includegraphics[width=\textwidth]{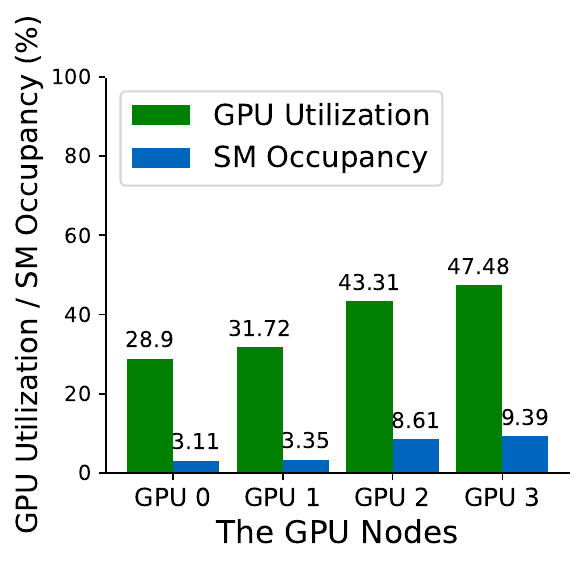}
    \vspace{-4mm}
    \caption{Only time sharing \cite{kubeshare_hpdc_2020}.}
    \label{fig:time_sharing_scheduling}
  \end{subfigure}
%   \quad % or any other spacing command you like
  \begin{subfigure}[b]{0.23\textwidth}
    \includegraphics[width=\textwidth]{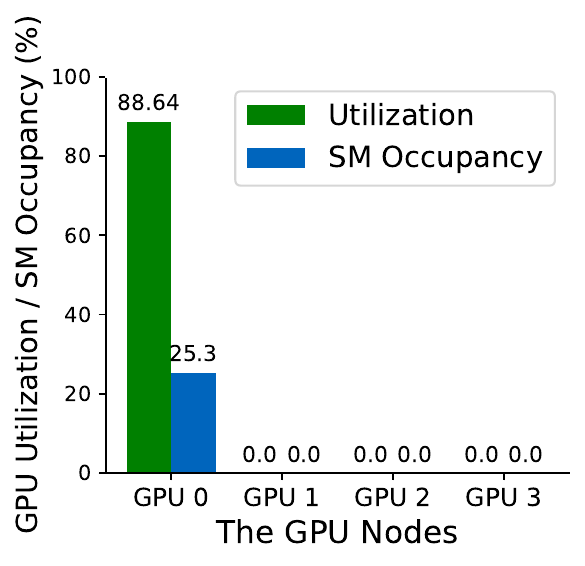}
    \vspace{-4mm}
    \caption{FaST-Scheduler.}
    \label{fig:fast_scheduler_result}
  \end{subfigure}
  \vspace{-2mm}
  \caption{The GPU utilization and SM occupancy under different scheduling mechanisms.}
  \label{fig:experiments_fast_scheduler}
  \vspace{-4mm}
\end{figure}

\begin{figure}[htbp]
\centering
\centering
\includegraphics[width=0.95\linewidth]{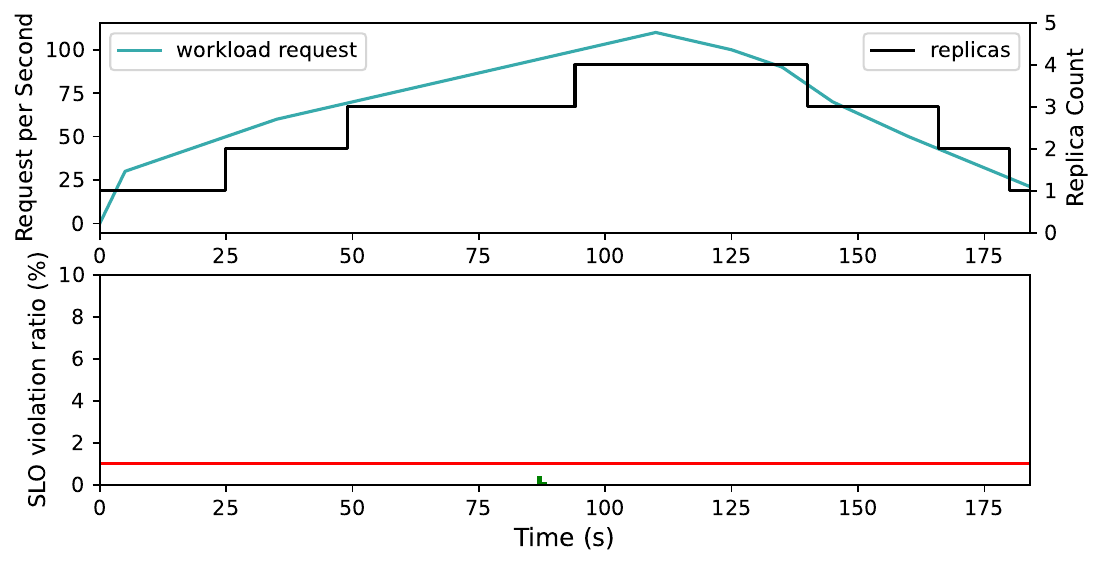}
\vspace{-4mm}
\caption{Auto-scaling to meet SLO.}
\label{fig:autoscaling_result}
\vspace{-3mm}
\end{figure}

To evaluate the efficiency of FaST-scheduler, we used workloads capable of scaling 4 ResNet pods, 2 RNNT pods, and 2 BERT pods, with (SM partitions, time quotas) set at (12\%, 40\%), (24\%, 40\%), and (50\%, 60\%), respectively. The experiment was conducted using 4 GPU work nodes, with scheduling carried out using both time sharing and FaST-Scheduler. The experiment results are shown in Figure \ref{fig:experiments_fast_scheduler}. The scheduling with only time sharing \cite{kubeshare_hpdc_2020} requires 4 GPUs to accommodate these function pods, each of them has low GPU utilization and SM occupancy.  However, the FaST-Scheduler, which is based on spatio-temporal sharing, only requires one GPU to handle these tasks. Compared to time sharing based scheduling, FaST-Scheduler is able to increase GPU utilization and SM occupancy by 1.34 times and 3.13 times, respectively. These results demonstrate the effectiveness and efficiency of the Maximal Rectangle Algorithm in significantly improving GPU usage. Meanwhile, as shown in Figure \ref{fig:autoscaling_result}, FaST-Scheduler can perform auto-scaling based on the current RPS to ensure function SLO. At the ResNet SLO of 69ms, the function did not experience SLO violations exceeding 1\%.

% . To assess the effectiveness of the FaST-scheduler, we created workloads capable of scaling 4 ResNet pods, 2 RNNT pods, and 2 BERT pods,

\begin{figure}[htbp]
  \centering
  \includegraphics[width=\linewidth]{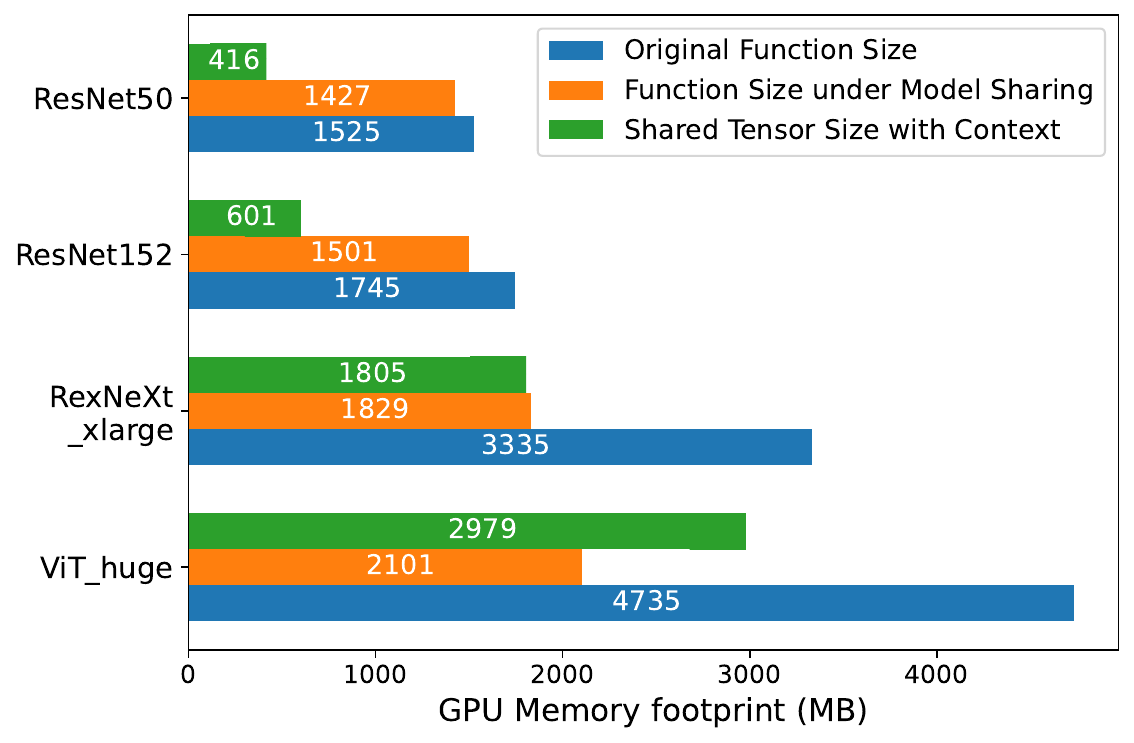}
  \vspace{-8mm}
  \caption{The GPU memory footprint of models.}
  \label{fig:model_sharing_result}
  \Description{}
  \vspace{-4mm}
\end{figure}
\subsection{Model Sharing}
To evaluate the effectiveness of model sharing in mitigating memory contention, we deployed models of various sizes and observed their GPU memory footprints, as illustrated in Figure \ref{fig:model_sharing_result}. The results show that compared to the original functions, all functions with model sharing reduced their memory footprint, with a more significant reduction observed for larger models. For example, the memory footprint of a single small ResNet model decreased by 6.4\% from 1525M to 1427M, and that of a large ViT\_huge model decreased by 55.6\% from 4735M to 2101M. This is because larger models have more model parameters. Meanwhile, The sharing introduces a fixed overhead of 300M in NVIDIA V100 GPU for each DL model to manage the storage process context, as shown in the hatched areas in Figure \ref{fig:model_sharing_result}. When deploying a function with a single pod instance, the GPU memory usage may be slightly higher. But for functions with multiple pod instances, the memory footprint is significantly reduced. For example, when deploying the ViT\_Huge function FaSTPod with only one pod, the memory footprint is 5080M with model sharing and 4735M without model sharing. However, if the same function contains 3 pods, the memory footprint is only 9282M with model sharing (2979M + 2101M x 3), compared to 14205M without model sharing (4735M x 3), resulting in a 4.8G reduction in memory footprint. Using model sharing, a 16G NVIDIA V100 GPU can accommodate 7 RexNeXt pods, whereas without model sharing, only 4 pods can be deployed. The experiment results demonstrate that model sharing can effectively mitigate GPU memory contention.

\vspace{-5pt}
\section{Conclusions}
This paper presents FaST-GShare, an efficient FaaS-oriented Spatio-Temporal GPU Sharing architecture for deep learning inferences. In FaST-GShare, we introduce the FaST-Manager to limit and isolate spatio-temporal resources for GPU multiplexing. In order to realize function performance, the automatic and flexible FaST-Profiler is proposed to profile function throughput under various resource allocation. Based on the profiling data and the isolation mechanism, we introduce the FaST-Scheduler with heuristic auto-scaling and efficient resource allocation to guarantee function SLOs. Meanwhile, FaST-Scheduler schedules functions with efficient GPU node selection to maximize GPU usage. Furthermore, model sharing is exploited to mitigate memory contention. Our prototype implementation on the OpenFaaS platform proves that FaST-GShare can ensure resource isolation and function SLOs. Compared to time sharing mechanism, FaST-GShare can improve throughput by 3.15x, GPU utilization by 1.34x and SM occupancy by 3.13x on average.

\vspace{-2mm}
\begin{acks}
This work is supported by the Google Cloud Research Credits Program. We sincerely thank Leibniz Supercomputing Centre (LRZ) for providing the GPU device support.
\end{acks}

\bibliographystyle{ACM-Reference-Format}
\bibliography{fast_share}

% \appendix









\end{document}